 \documentclass[final,5p,times,twocolumn,authoryear]{elsarticle}

\usepackage{amssymb}
\usepackage{lipsum}

\usepackage{subfig}

\usepackage[final]{changes}

\begin{document}

\begin{frontmatter}



\title{Image of the time-dependent black hole}


\author[first]{Yu Liang}
\ead{washy2718@outlook.com}

\author[second]{Sen Guo}
\ead{sguophys@126.com}

\author[third]{Kai Lin}
\ead{lk314159@hotmail.com}

\author[fourth]{Yu-Hao Cui}

\author[fifth]{Yu-Xiang Huang}

\author[first]{Tao Yu}

\affiliation[first]{organization={China University of Geosciences},
            city={Wuhan},
            postcode={430074}, 
            country={China}}
\affiliation[second]{organization={Chongqing Normal University},
            city={Chongqing},
            postcode={401331}, 
            country={China}}
\affiliation[third]{organization={Universidade Federal de Campina Grande},
            city={Campina Grande},
            postcode={PB}, 
            country={Brasil}}
\affiliation[fourth]{organization={Xiamen University},
            city={Xiamen},
            postcode={361005}, 
            country={China}}
\affiliation[fifth]{organization={Guangxi University},
            city={Nanning},
            postcode={530004}, 
            country={China}}

\begin{abstract}
The Event Horizon Telescope’s 2024 observation\textcolor{black}{s} report a shift in the position angle of the brightness asymmetry in M87*, revealing time variability in the black hole’s image. 
In this analysis, we investigate the time-dependent of a Vaidya black hole. 
By introducing a mass function that increases linearly with time, \textcolor{black}{along with} a conformal transformation, we \textcolor{black}{derive} the conformal Vaidya metric and define a new time coordinate $t_c$. 
\textcolor{black}{Using} the semi-analytical approach, we \textcolor{black}{analyze} the ray trajectories and radiation flux of the \textcolor{black}{Vaidya} black hole \textcolor{black}{in the background of a thin accretion disk}. 
We discuss how the observed flux in the Vaidya spacetime evolves as a function of \textcolor{black}{the new time coordinate} $t_c$. 
The results show that the facula on the observable plane undergoes radial displacement as $t_c$ increases, revealing the time-dependent evolution of black hole images.
\end{abstract}




\end{frontmatter}




\section{Introduction}
\label{introduction}

In 2024, the Event Horizon Telescope (EHT) released a new set of images of the M87* black hole, based on data collected in 2018, and compared them with the 2017 observations. 
The results indicated that the position angle of \textcolor{black}{the} brightness asymmetry shifted from approximately 180$^\circ$ to around 215$^\circ$, corresponding to a counterclockwise rotation of about 35$^\circ$ \citep{EventHorizonTelescope:2024dhe}. 
\textcolor{black}{This} dynamic change in angular position has drawn attention to the time-dependent evolution of black hole images. 
\textcolor{black}{Focusing} solely on the spacetime near the black hole and \textcolor{black}{neglecting external effects} during light propagation, two potential mechanisms may \textcolor{black}{contribute} to \textcolor{black}{the time evolution of} black hole images. 
One possibility is the anisotropy and motion of the surrounding radiation sources, while the other \textcolor{black}{involves} the evolution of the spacetime background driven by changes in the black hole’s mass. 
The first hypothesis can be further refined by considering the origin and structure of the emission sources, such as hotspot\textcolor{black}{s}, dynamic accretion disk models, and so on \citep{Huang:2024wpj,Lee:2020pvs}. 
Previously, Mishra et al.\ derived the evolution equation for the photon sphere in dynamical black hole spacetimes and \textcolor{black}{investigated} its behavior in \textcolor{black}{various} black hole models through numerical solutions \citep{Mishra:2019trb}. 
Solanki and Perlick \textcolor{black}{investigated} the dynamical evolution of both the photon sphere and black hole shadow in the context of the Vaidya metric, \textcolor{black}{while} Tan \textcolor{black}{analyzed} the evolution of black hole shadows in a Kerr-Vaidya-like \textcolor{black}{spacetime} \citep{Solanki:2022glc,Tan:2023ngk}. 

The EHT has successively released images of the black holes at the centers of the M87 galaxy and the Milky Way, providing observational confirmation of the black hole shadow\textcolor{black}{s} \citep{EventHorizonTelescope:2019dse,EventHorizonTelescope:2022wkp}. 
\textcolor{black}{These observations have} sparked widespread interest among researchers in the optical appearance of black holes. 
\textcolor{black}{In 1979,} Luminet first presented the optical appearance of a Schwarzschild black hole using a semi-analytical method \citep{Luminet:1979nyg}. Since then, \textcolor{black}{numerous} studies on black hole shadows have emerged. 
In 2019, Gralla et al.\ classified light rays into direct, lensed, and photon ring components by defining the number of orbits, focusing on Schwarzschild and Kerr black holes surrounded by optically and geometrically thin disks \citep{Gralla:2019xty,Gralla:2019drh}. 
\textcolor{black}{Additionally}, Gyulchev et al.\ studied the optical appearance and radiation properties of thin accretion disks around the static Janis–Newman–Winicour naked singularity, while Gan et al.\ explored the visual appearance of a class of hairy black holes \citep{Gyulchev:2019tvk, Gan:2021pwu}. 
In 2022, Perlick and Tsupko conducted a comprehensive review of the current state of research on black hole shadows\citep{Perlick:2021aok}. 
In recent years, similar research has also appeared in the works of Peng et al.\ and other researchers \citep{Peng:2020wun,Peng:2021osd,Hou:2022eev,Zeng:2022pvb,Zeng:2023zlf,Wang:2023vcv,Huang:2023ilm,Meng:2023htc,Meng:2024puu,Cui:2024wvz,Guo:2022iiy,Guo:2024mij}. 
The approaches adopted in these \textcolor{black}{studies} have significantly contributed to the smooth advancement of this research.

However, these studies did not further \textcolor{black}{investigate} photon trajectories and radiation flux within a \textcolor{black}{dynamical} spacetime background, which constitutes the core \textcolor{black}{focus} of this \textcolor{black}{work}.
In this work, we model the black hole as a Vaidya solution with a linearly increasing mass function and derive its evolving image when surrounded by an accretion disk, \textcolor{black}{thereby} providing a theoretical framework \textcolor{black}{for understanding the} observational signatures of temporal variability. 
The structure of \textcolor{black}{this paper} is as follows: we perform a conformal transformation of the Vaidya metric and solve the null geodesic equations in the conformal Vaidya spacetime in Section 2. 
Section 3 discuss\textcolor{black}{es} the observer’s plane in the conformal Vaidya spacetime and use\textcolor{black}{s} ray-tracing to calculate both the direct and secondary images. 
Section 4 evaluate\textcolor{black}{s} the observed flux of a black hole surrounded by a thin accretion disk within the conformal spacetime. 
Finally, we explore the temporal behavior of the observed flux for the Vaidya black hole in the $t_c-r$ coordinate system and \textcolor{black}{summarize the main findings}.

\section{Null Geodesics of the Vaidya black hole}

For a Vaidya black hole with a mass increasing over time, the metric in the ingoing Eddington–Finkelstein coordinate system \textcolor{black}{is given by} the following form
\begin{equation}\label{eq:c_ds2_v}
    ds^2 = -f dv^2 + 2dvdr + r^2 d\theta^2 + r^2 \sin^2 \theta d\phi^2
\end{equation}
where $f = 1 - 2 m(v) / r$. In the special case where the mass function $m(v)$ \textcolor{black}{increases linearly with $v$}, the metric in Eq.~(\ref{eq:c_ds2_v}) admits an additional conformal Killing vector\citep{Ojako:2019gwc,Solanki:2022glc}. Let $m(v) = \mu v$, where $\mu$ is a positive constant. Then we have
\begin{equation}
    f = 1 - \frac{2 \mu v}{r}
\end{equation}

To make the conformal Killing vector explicit, we define the conformal transformation factor as follows
\begin{equation}
    \Theta = e^{\nu_c / r_0}
\end{equation}
where $r_0$ is a positive constant. \textcolor{black}{We then} perform the following conformal transformation \citep{Solanki:2022glc}
\begin{equation} \label{eq:coord_trans}
    v = r_0 \Theta, \quad r = R \Theta.
\end{equation}
After simplification, the metric in the conformal ingoing Eddington–Finkelstein coordinate system takes the \textcolor{black}{following} form
\begin{equation}
    ds^2 = \Theta^2 \left( -f_c dv_c^2 + 2 dv_c dR + R^2 d\theta^2 + R^2 \sin^2 \theta d\phi^2 \right)
\end{equation}
where
\begin{equation}
    f_c = 1 - \frac{2 \mu r_0}{R} - \frac{2R}{r_0}
\end{equation}
It can be seen that $\partial / \partial \nu_c$ is defined as a conformal Killing vector. Moreover, since the metric is conformal to a static spacetime, it can be taken to generate a rotating solution via the Newman–Janis algorithm. At this point, it is necessary to introduce a \textcolor{black}{new} time coordinate $t_c$, defined \textcolor{black}{as follows}
\begin{equation}
    v_c = t_c + F_c(R), \quad F_c(R) \equiv \int \frac{dR}{f_c}.
\end{equation}
After simplification and \textcolor{black}{coordinate transformation}, the Vaidya metric in the conformal spherical coordinate system \textcolor{black}{takes the following form}
\begin{equation} \label{eq:c_ds2}
    d s^2 = \Theta^2 \left( -f_c dt_c^2 + \frac{dR^2}{f_c} + R^2 d\theta^2 + R^2 \sin^2 \theta d\phi^2 \right)
\end{equation}
To ensure that the conformal spacetime \textcolor{black}{remains} timelike, we require $f_c > 0$, which \textcolor{black}{implies that} the conformal Killing horizon \textcolor{black}{is} located at
\begin{equation}\label{eq:c_Rpm}
    R_\pm = \frac{r_0}{4} (1 \pm \sqrt{1-16\mu}), \quad \mu < 1/16.
\end{equation}
\textcolor{black}{where $R_+$ and $R_-$ denote the outer and inner conformal Killing horizons, respectively.} Accordingly, we have $-\infty < t_c < +\infty$ and $R_- < R < R_+$.

To simplify the analysis, \textcolor{black}{we restrict our attention to null geodesics confined to} the equatorial plane. \textcolor{black}{Based on} the metric~(\ref{eq:c_ds2}), the \textcolor{black}{corresponding} Lagrangian is given by
\begin{equation}
    \widetilde{L} = \frac{\Theta^2}{2} \left[ - f_c \dot{t_c}^2 + \frac{\dot{R}^2}{f_c} + R^2 \dot{\phi}^2 \right] = 0
\end{equation}
\textcolor{black}{Since} $\partial / \partial t_c$ is also a conformal Killing vector, both the photon energy $E$ and angular momentum $L$ \textcolor{black}{are conserved} \citep{Solanki:2022glc}. These \textcolor{black}{conserved} quantities are given by
\begin{equation}
    \frac{\partial \widetilde{L}}{\partial \dot{t_c}} = - \Theta^2 f_c \dot{t_c} \equiv E
\end{equation}
\begin{equation}
    \frac{\partial \widetilde{L}}{\partial \dot{\phi}} = \Theta^2 R^2 \dot{\phi} \equiv L
\end{equation}
When $\dot{\phi} \neq 0$, the above three equations \textcolor{black}{lead to} the null geodesic equation in the conformal Vaidya spacetime, \textcolor{black}{given by}
\begin{equation} \label{eq:c_pVE}
    \left[ \frac{1}{R^2} \left( \frac{d R}{d \phi}\right) \right]^2 + \frac{f_c}{R^2} = \frac{1}{b^2}
\end{equation}
where $b = L/E$ is the impact parameter. \textcolor{black}{It follows} that the photon trajectory in the conformal coordinate system is independent of the time coordinate $t_c$.

For \textcolor{black}{light rays} that are not captured by a black hole, there exists a periastron $R_m$ \textcolor{black}{along the} trajectory such that $d R_m / d \phi = 0$. From Eq.~(\ref{eq:c_pVE}), the relation\textcolor{black}{ship} between the impact parameter and the periastron is given by
\begin{equation} \label{eq:b_Rm}
    b = \frac{R_m}{\sqrt{f_c}} = \sqrt{\frac{R_m^3}{R_m - 2 \mu r_0 - 2 R_m^2 / r_0}}
\end{equation}
Since the radius of the photon sphere $R_p$ \textcolor{black}{corresponds to} the minimum among all periastra, it must also satisfy \textcolor{black}{the condition} $d^2 R/d \phi^2 = 0$. Solving this condition yields \citep{Solanki:2022glc}
\begin{equation}
    R_p = \frac{r_0}{2} (1 - \sqrt{1 - 12\mu})
\end{equation}
By substituting $R_p$ into Eq.~(\ref{eq:b_Rm}), the critical impact parameter $b_p$ \textcolor{black}{can be determined}. When $b>b_p$, \textcolor{black}{light rays are} deflected but not captured by the black hole, \textcolor{black}{whereas for} $b<b_p$, \textcolor{black}{they are} captured.

\section{Ray trajectory of the Vaidya black hole}

\begin{figure*}
    \centering
    \subfloat{\includegraphics[width=.35\textwidth]{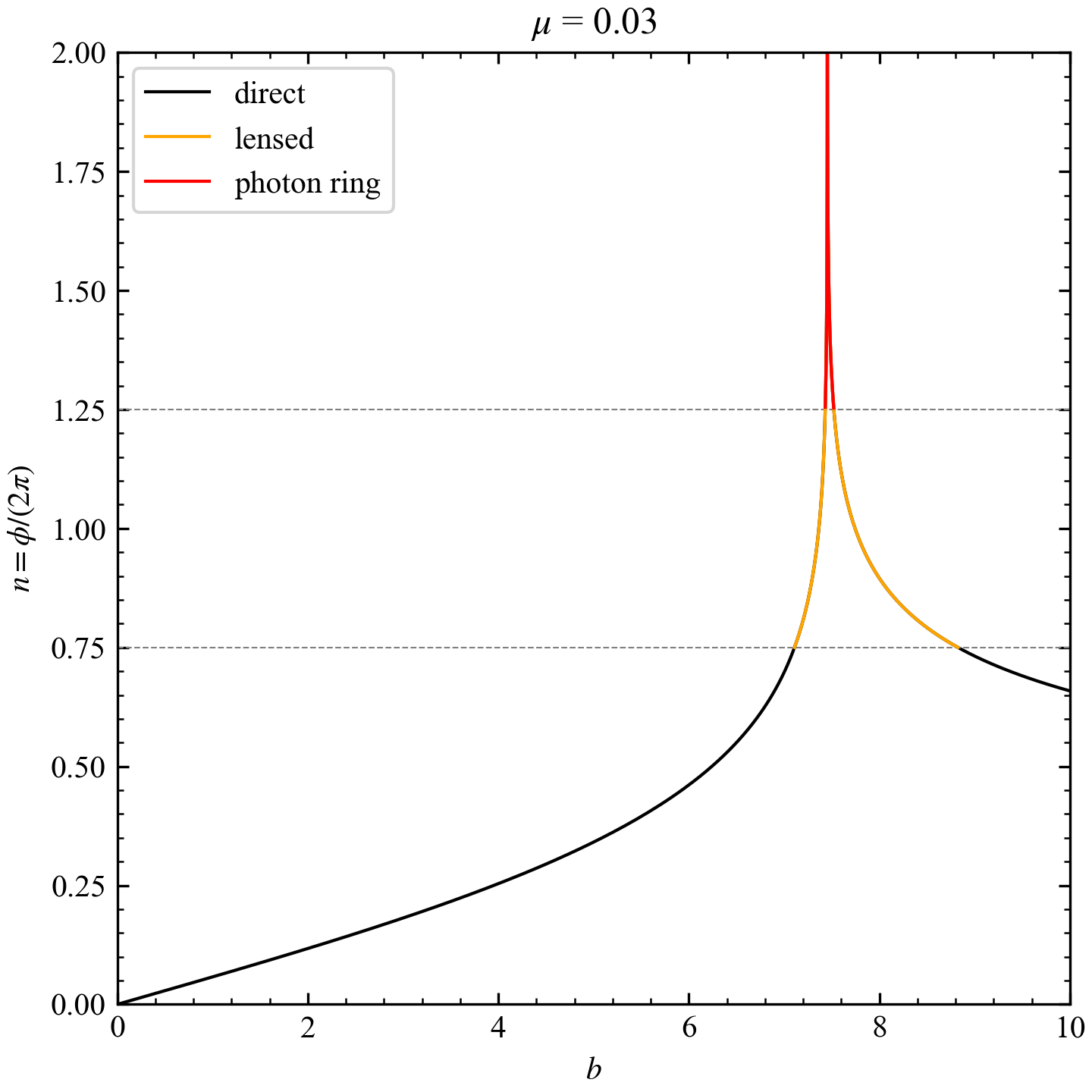}}\hspace{.1\textwidth}
    \subfloat{\includegraphics[width=.35\textwidth]{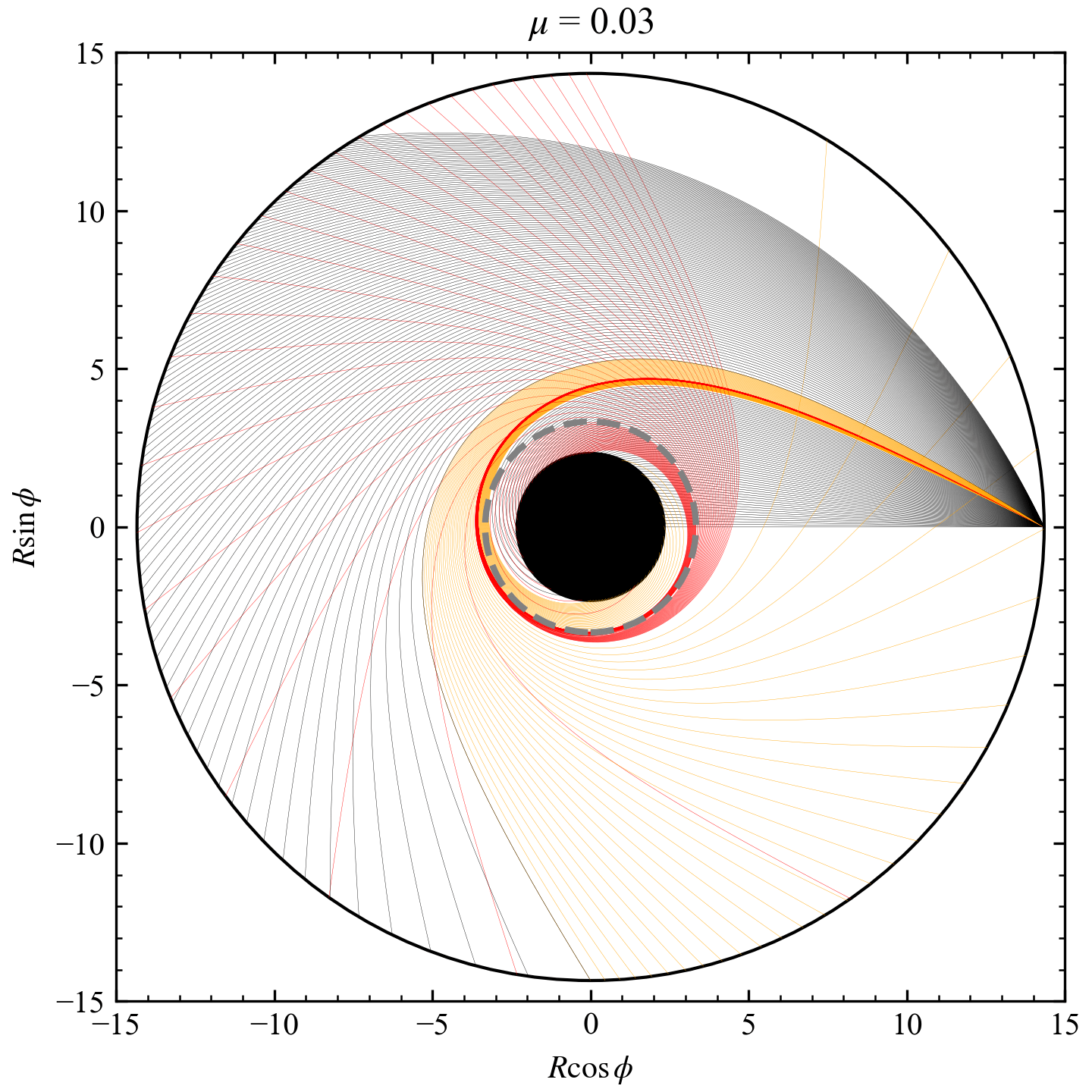}}
    \caption{\label{fig:b_gamma}Behavior of photons in the conformal Vaidya spacetime as a function of impact parameter $b$. \textcolor{black}{The left panel illustrates} the dependence of the orbital number $n$ is illustrated, with direct, lensed, and photon ring components marked in black, orange, and red, respectively. The right \textcolor{black}{panel shows} the corresponding photon trajectories.}
\end{figure*}

\begin{figure*}
    \centering
    \includegraphics[width=.45\textwidth]{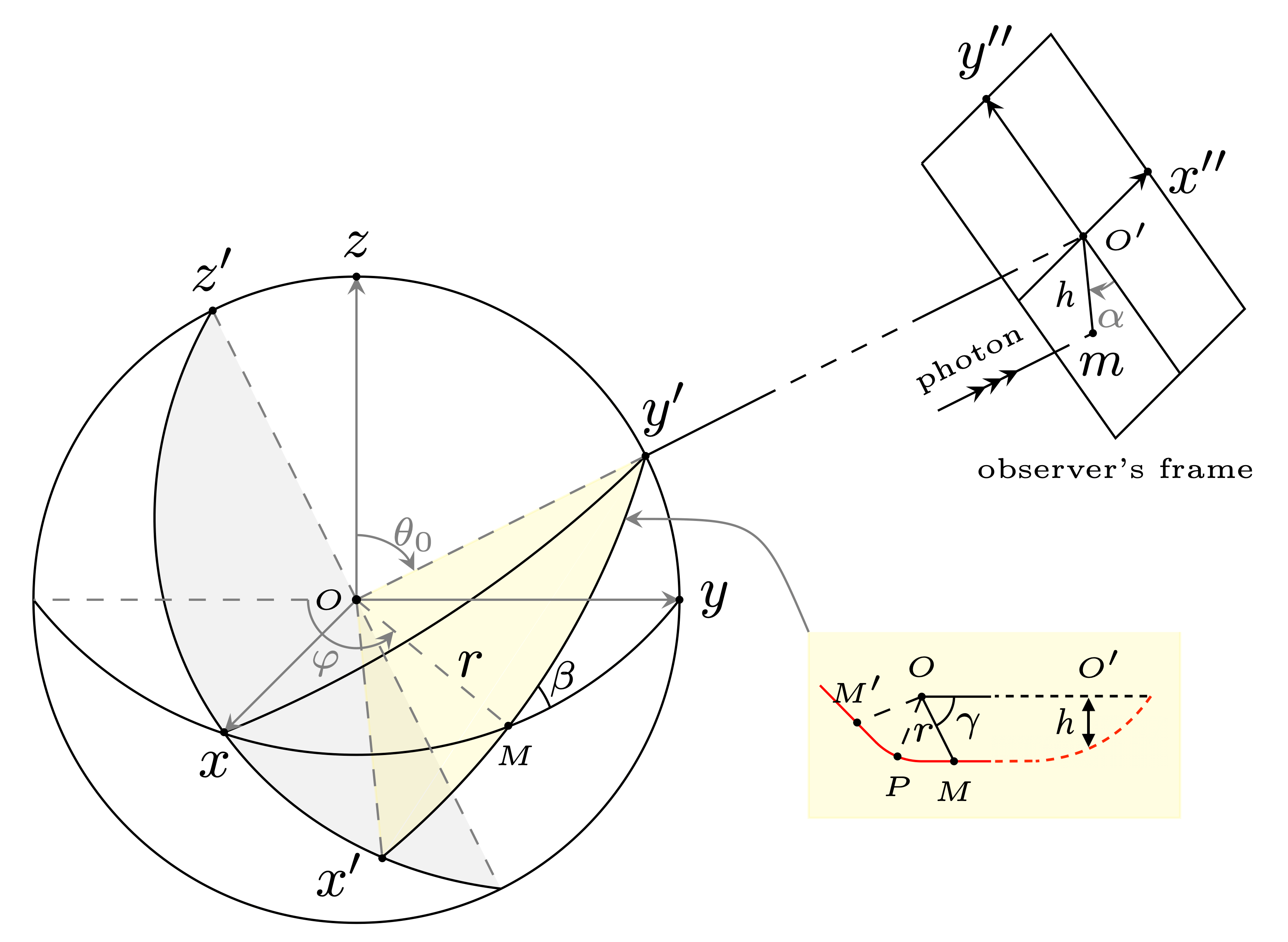}
    \includegraphics[width=.45\textwidth]{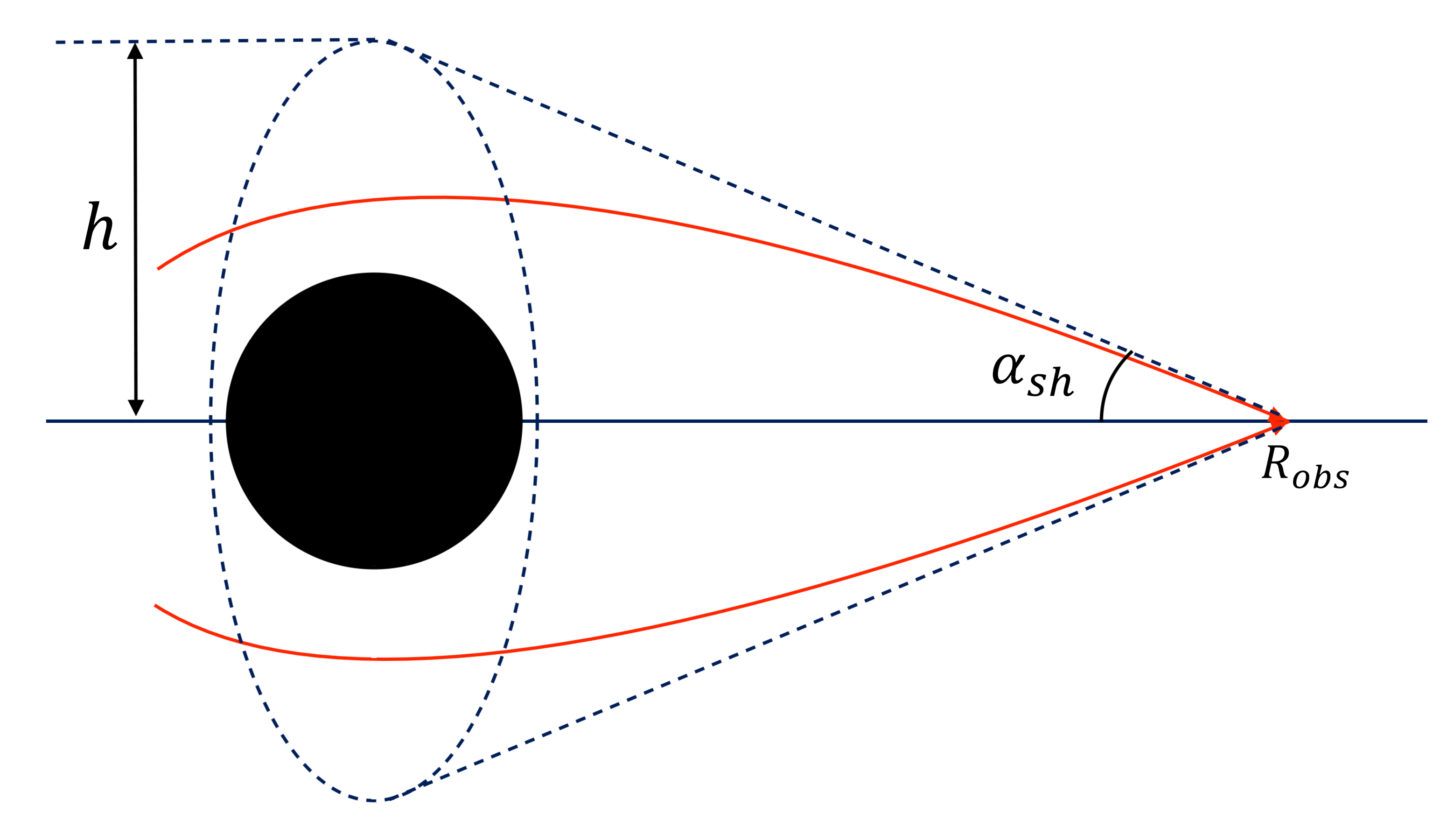}
    \caption{\label{fig:coordinate} The conformal coordinate system. The left \textcolor{black}{panel} follows the coordinate \textcolor{black}{framework introduced by} Luminet and Tian \citep{Luminet:1979nyg,Tian:2019yhn}, with two \textcolor{black}{key modifications}: (1) the projection distance on the observer’s plane is denoted by $h$ instead of the impact parameter $b$; (2) light rays \textcolor{black}{are assumed to} converge at $R_+$ on the yellow plane. The right \textcolor{black}{panel} illustrates the geometric relation among the observer’s position $R_{obs}$, the angular radius $\alpha_{sh}$, and \textcolor{black}{the projection distance} $h$.}
\end{figure*}

\begin{figure*}
    \centering
    \subfloat{\includegraphics[width=.3\textwidth]{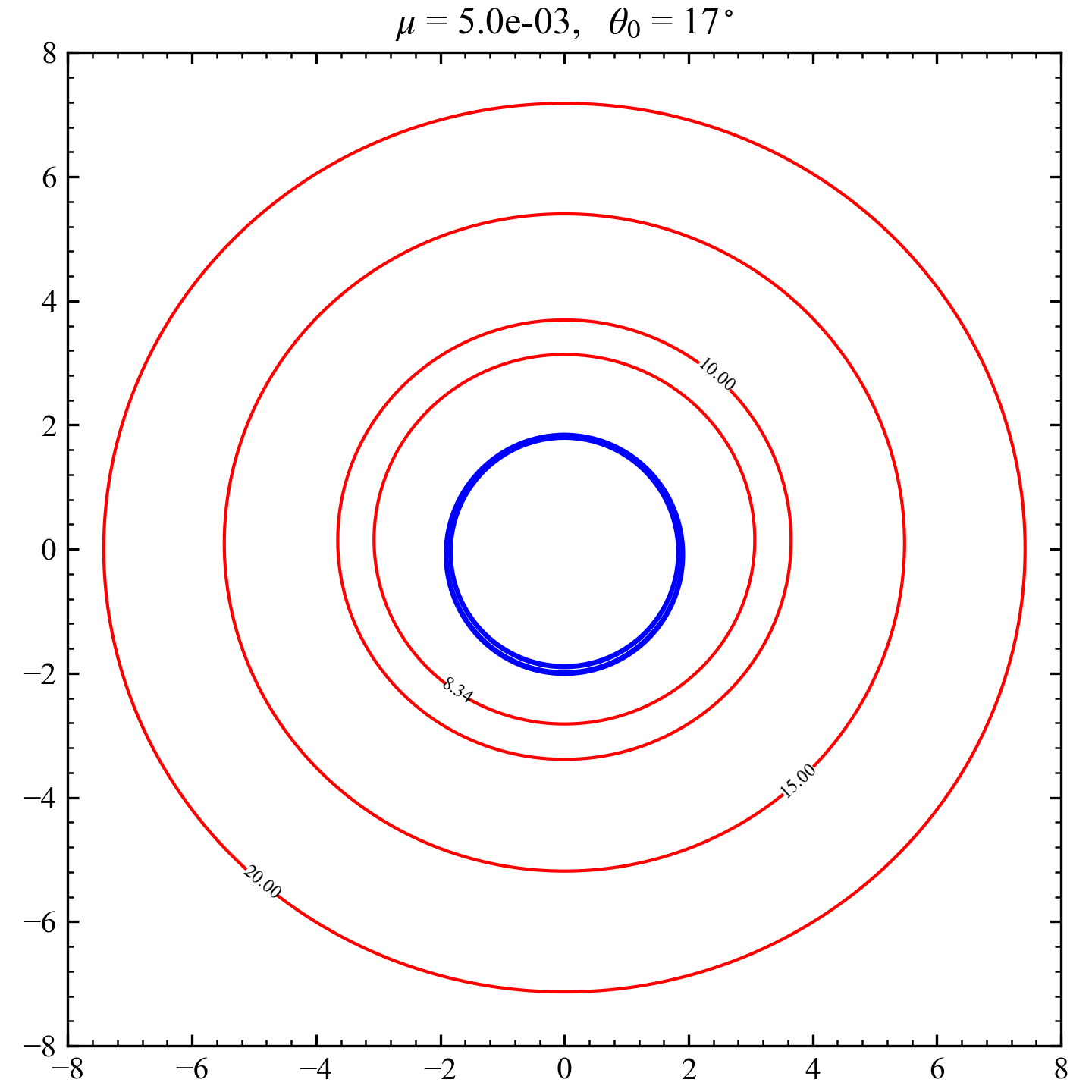}}
    \subfloat{\includegraphics[width=.3\textwidth]{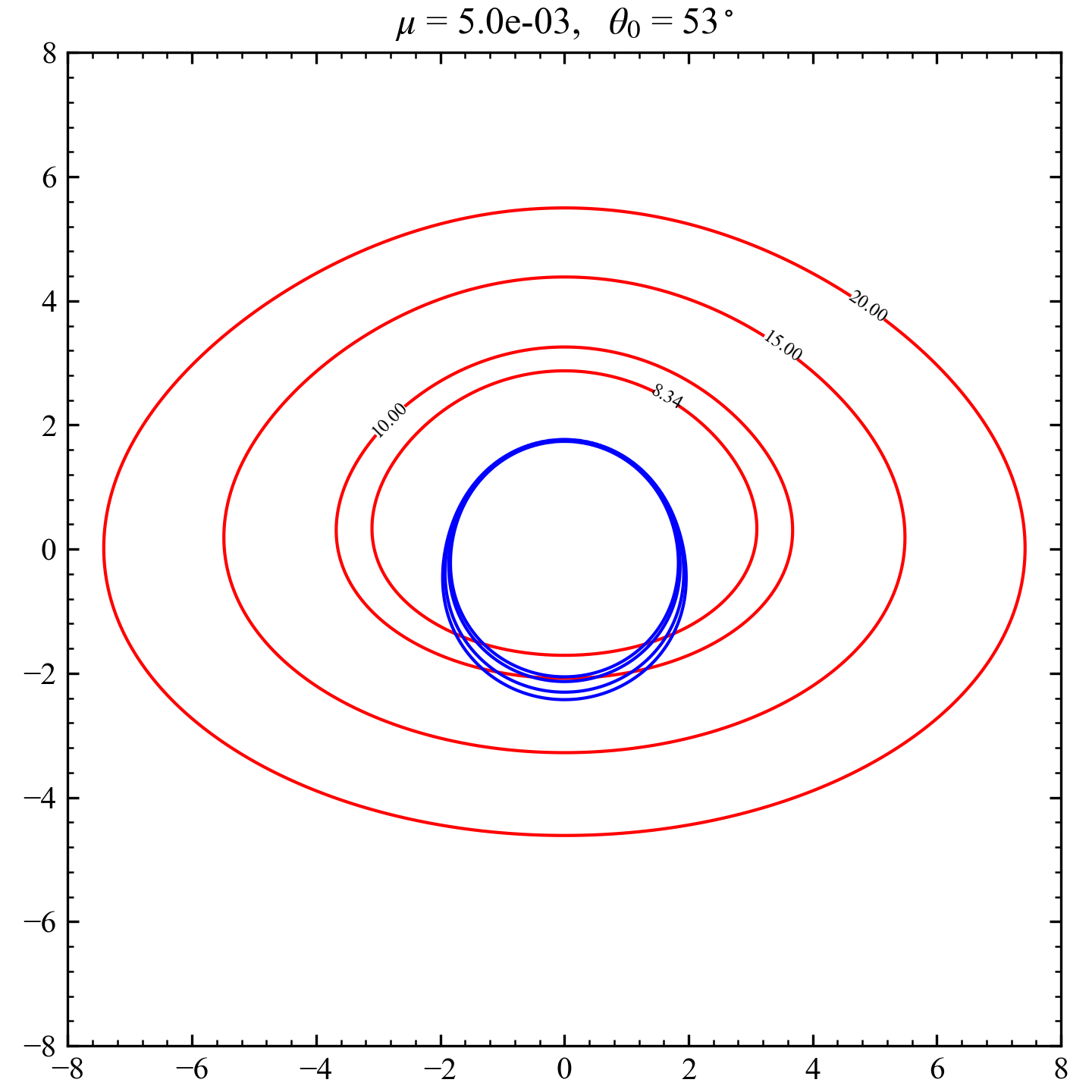}}
    \subfloat{\includegraphics[width=.3\textwidth]{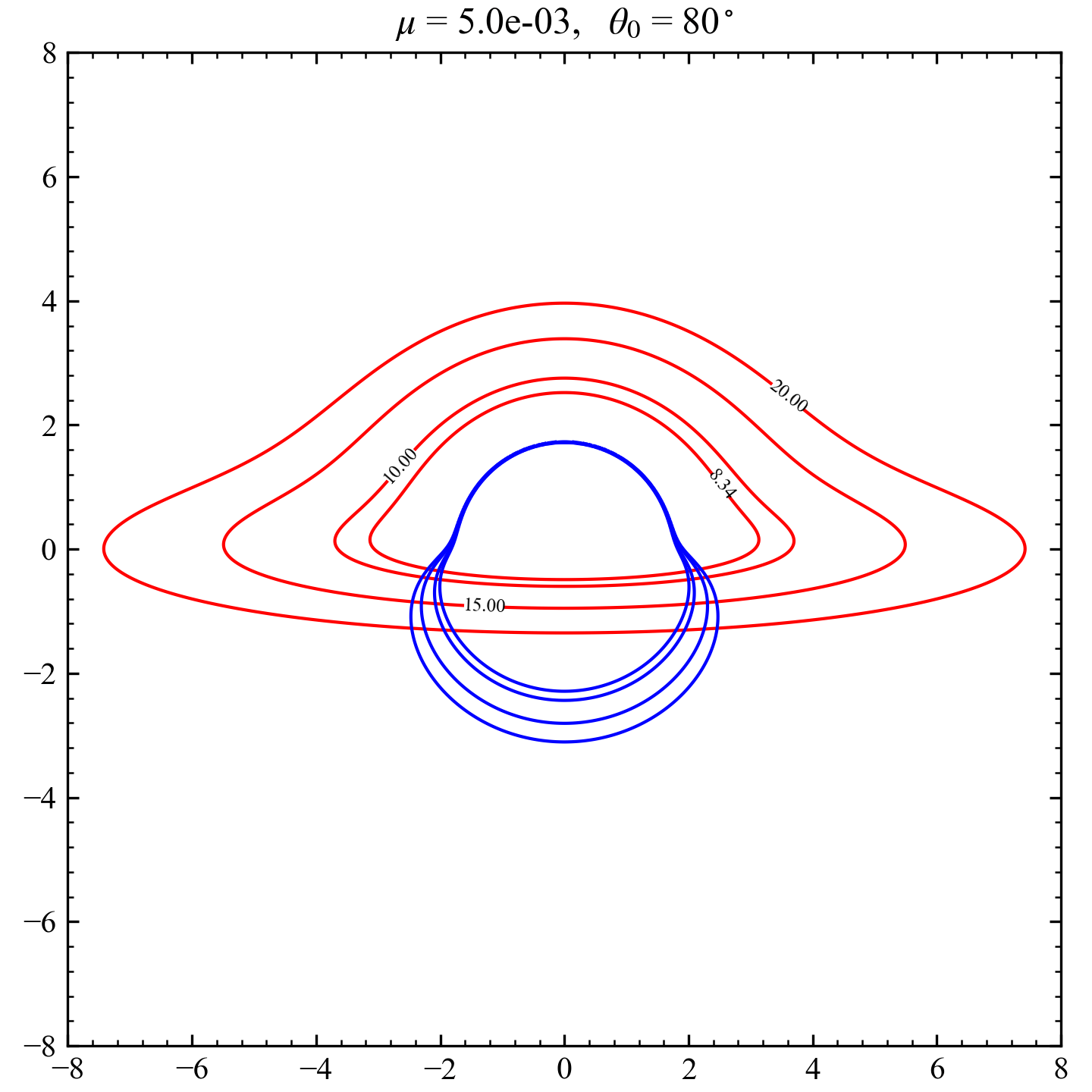}} \\
    \subfloat{\includegraphics[width=.3\textwidth]{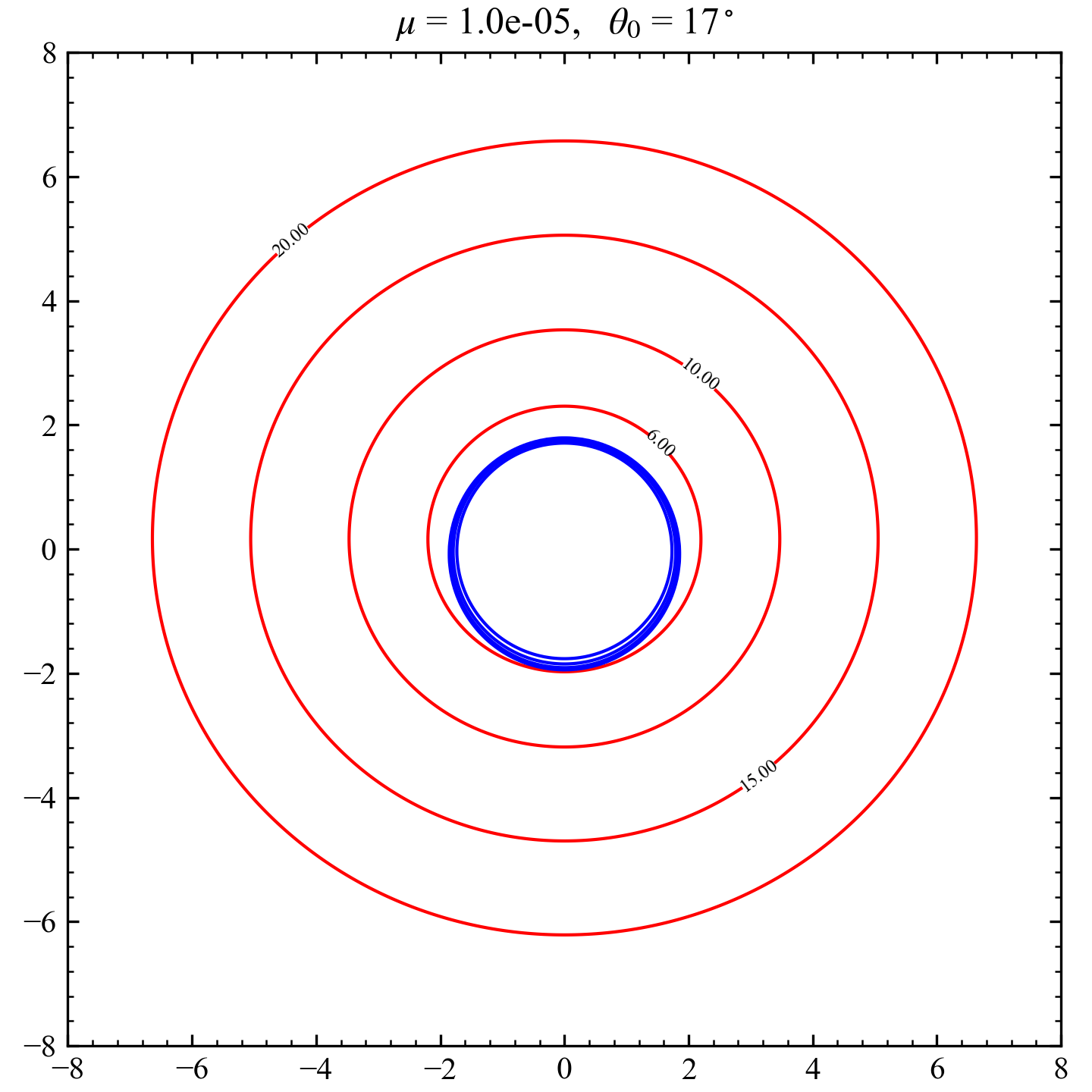}}
    \subfloat{\includegraphics[width=.3\textwidth]{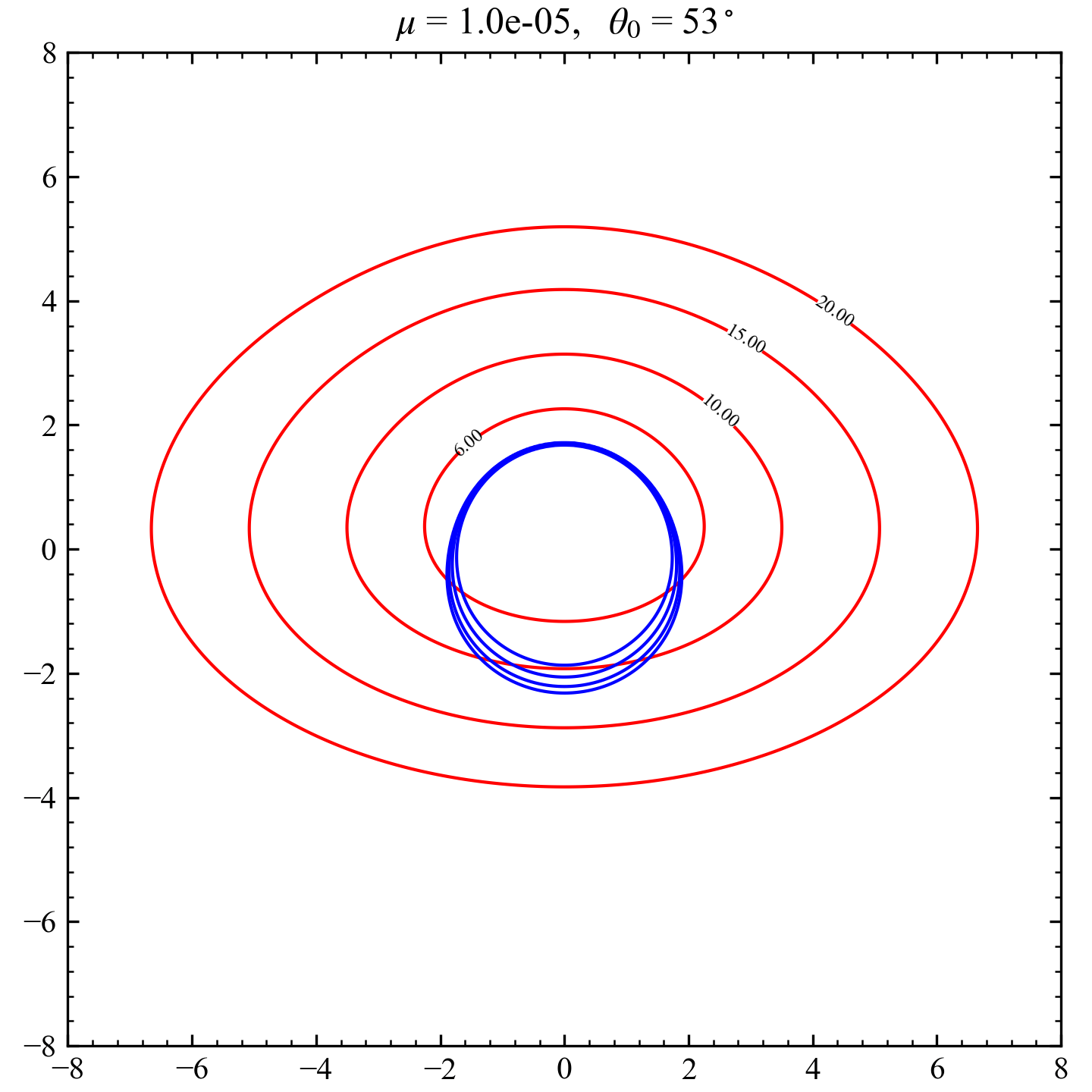}}
    \subfloat{\includegraphics[width=.3\textwidth]{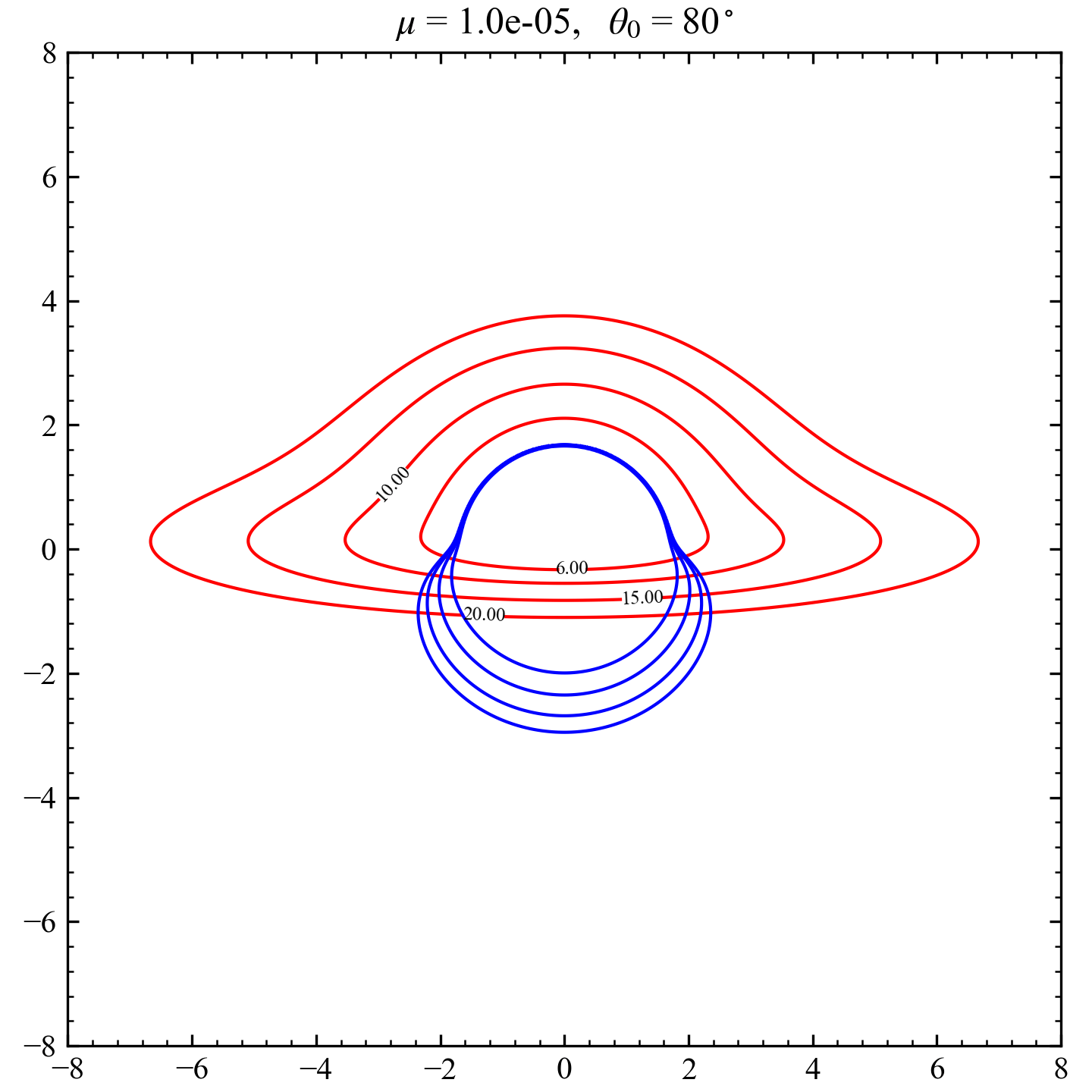}}
    \caption{\label{fig:c_images}\textcolor{black}{The direct (red) and secondary (blue) images} of a thin accretion disk in the conformal Vaidya spacetime background.}
\end{figure*}

\subsection{The deflection angle of the light ray}

For a Schwarzschild black hole, light rays are typically assumed to originate from or propagate to spatial infinity. \textcolor{black}{In contrast}, for the conformal Vaidya black hole, \textcolor{black}{the radial coordinate $R$ is bounded} between $R_-$ and $R_+$, \textcolor{black}{implying that light rays must be considered as originating from or terminating at} $R_+$. By defining $R = 1 / U$, Eq.~(\ref{eq:c_pVE}) can be \textcolor{black}{recast} as
\begin{equation}
    \left( \frac{dU}{d\phi} \right)^2 = 2 M_0 U^3 - U^2 + \frac{2U}{r_0} + \frac{1}{b^2} \equiv G(U)
\end{equation}
where $M_0 \equiv \mu r_0$ is a constant. Throughout this paper, we \textcolor{black}{set} $M_0 = 1$ for all calculations.

Although the above expression \textcolor{black}{gives the} differential form of $\phi$, its integral form is required to accurately \textcolor{black}{compute} the deflection angle. For a light ray emitted from $R_+$, if $b < b_p$, or if $b > b_p$ but the ray does not reach the periastron $R_m$, the deflection angle is \textcolor{black}{given by} the following integral expression
\begin{equation} \label{eq:phi_0}
    \phi = \int_{U_+}^{U} \frac{d U}{\sqrt{G(U)}}
\end{equation}
If $b > b_p$ and the photon has already passed through \textcolor{black}{the periastron} $R_m$, then the deflection angle \textcolor{black}{is given by}
\begin{equation} \label{eq:phi_1}
    \phi = 2 \int_{U_+}^{U_m} \frac{d U}{\sqrt{G(U)}} - \int_{U_+}^{U} \frac{d U}{\sqrt{G(U)}}
\end{equation}
where $U_+ = 1 / R_+$ and $U_m = 1 / R_m$.

\textcolor{black}{We} define the orbit number as $n \equiv \phi / 2\pi$, where $\phi$ is the total azimuthal change outside the horizon. When $n < 3/4$, the photon crosses the equatorial plane only once; when $n > 3/4$, it crosses at least twice; and when $n > 5/4$, it crosses at least three times\citep{Gralla:2019xty}. The relation\textcolor{black}{ship} between the orbit number $n$ and the impact parameter $b$, as well as the corresponding light trajectories, \textcolor{black}{is} shown in Fig.\ref{fig:b_gamma} for $\mu = 0.03$. 
As \textcolor{black}{illustrated}, the photon trajectories no longer \textcolor{black}{extend to spatial} infinity but instead converge at $R_+$. 
According to Eq.~(\ref{eq:c_Rpm}), $R_+$ increases as $\mu$ decreases. In the limit $\mu \to 0$, $R_+ \to +\infty$, and the photon trajectories \textcolor{black}{approach} those of a Schwarzschild black hole\citep{Tan:2023ngk}.

\subsection{Ray trajectory}

\textcolor{black}{We} consider a Vaidya black hole surrounded by a geometrically thin accretion disk. 
Assume that the observer’s image plane is \textcolor{black}{positioned} at $R_{obs}$ and inclined at an angle $\theta_0$ \textcolor{black}{relative} to the equatorial plane. 
A conformal coordinate system is \textcolor{black}{introduced, as illustrated} in Fig.~\ref{fig:coordinate}. 
The construction of Fig.\ref{fig:coordinate} follows the \textcolor{black}{framework established by} Luminet and Tian \citep{Luminet:1979nyg,Tian:2019yhn}, \textcolor{black}{with two key modifications}. 
First, the distance on the observer’s plane is \textcolor{black}{described} by a new parameter $h$, \textcolor{black}{which is} determined by the observer’s position and the angular radius, rather than \textcolor{black}{by} the impact parameter $b$. 
Second, the light rays \textcolor{black}{are assumed to} converge at $R_+$. 
\textcolor{black}{Despite these changes}, the original angular \textcolor{black}{relationships} remain \textcolor{black}{valid} and \textcolor{black}{continue to} satisfy the following expressions
\begin{equation}
    \cos \gamma = \frac{\cos{\alpha}}{\sqrt{\cos^2{\alpha} + \cot^2{\theta_0}}}
\end{equation}

When \textcolor{black}{specifying} the position of the observation plane, \textcolor{black}{it is important to} note that $R_{obs}$ \textcolor{black}{must lie strictly} between the two horizons and cannot \textcolor{black}{coincide with} either horizon. 
Based on the geometric relation\textcolor{black}{ship}s \textcolor{black}{depicted} in Fig.~\ref{fig:coordinate}, we obtain
\begin{equation} \label{eq:hc}
    h = R_{obs} \tan \alpha_{sh}
\end{equation}
where $\alpha_{sh}$ is the angular radius. 
\textcolor{black}{For a light ray with impact parameter $b$, the corresponding angular radius $\alpha_{sh}$ satisfies the following relation} \citep{Solanki:2022glc}
\begin{equation}
    \tan \alpha_{sh} = \sqrt{\frac{f_c(R_{obs})}{R_{obs}^2 / b^2 - f_c(R_{obs})}}
\end{equation}

For a light source located at $R$, the deflection angle $\phi = \gamma$ for the direct image can be computed using Eq.~(\ref{eq:phi_0}) or Eq.~(\ref{eq:phi_1}). For the $(1+n)$th-order image, the deflection angle is given by $\phi = 2n\pi - \gamma$, and can be evaluated using Eq.~(\ref{eq:phi_1}). In practice, by substituting known parameters ($\phi$, $R$) and reformulating the equation as a root-finding problem, the unknown impact parameter $b$ can be efficiently determined using Python or other numerical solvers.

To determine the inner edge of the accretion disk, \textcolor{black}{it is necessary} to compute the innermost stable circular orbit (ISCO). 
For the \textcolor{black}{spacetime described by the metric} in Eq.~(\ref{eq:c_ds2}), \textcolor{black}{the conditions for a circular orbit are $\dot{r} = 0$ and $\ddot{r} = 0$. Imposing these conditions yields}
\begin{equation}
    E^2 = \frac{2 \Theta^2 \left( f_c^2 + R/r_0 f_c \right)}{2 f_c - R f_c'}, \label{eq:c_E2_p}
\end{equation}
\begin{equation}
    L^2 = \frac{ \Theta^2 R^3 (f_c' + 2/r_0)}{2 f_c - R f_c'}.\label{eq:c_L2_p}
\end{equation}
\textcolor{black}{Given} the Hamiltonian of the test particle $\widetilde{H} = -1/2$, \textcolor{black}{the corresponding effective potential is expressed as}
\begin{equation}\label{eq:c_V_p}
    V_m(R) = f_c \left( \frac{L^2}{R^2} + \Theta^2 \right)
\end{equation}
\textcolor{black}{By} combining the equation $V "(R) = 0$ \textcolor{black}{with} the three equations \textcolor{black}{above}, \textcolor{black}{the ISCO radius is found to satisfy the following equation}
\begin{equation} \label{eq:ISCO}
    6 R^4 - 8 r_0 R^3 + (1 + 22 \mu) r_0^2 R^2 - 8 \mu r_0^3 R + 12 \mu^2 r_0^4 = 0
\end{equation}
An analysis of Eq.~(\ref{eq:ISCO}) \textcolor{black}{reveals} that for $\mu < 6.063 \times 10^{-3}$, there \textcolor{black}{exist} four real roots, \textcolor{black}{with} the ISCO corresponds to the second smallest among them. 
\textcolor{black}{For} $\mu > 6.063 \times 10^{-3}$, no ISCO exists within the physically \textcolor{black}{admissible} region. 
\textcolor{black}{Moreover, the ISCO radius increases monotonically with $\mu$, attaining its minimum value of 6 at $\mu = 0$, and reaching a maximum of approximately $R_{ISCO} \approx 9.421$ as $\mu$ approaches $6.063 \times 10^{-3}$.}

Figure~\ref{fig:c_images} \textcolor{black}{displays} the direct (red) and secondary (blue) images of a thin accretion disk in the conformal Vaidya spacetime for $\mu = 5 \times 10^{-3}$ and $10^{-5}$. 
\textcolor{black}{In both cases}, the \textcolor{black}{observer is located at a radial position of} $0.9 R_+$. 
\textcolor{black}{As shown in the figure,} for a fixed value of $\mu$, both the direct and secondary images are distorted as the inclination angle varies. 
\textcolor{black}{Conversely,} for a fixed inclination angle, the image size \textcolor{black}{decreases with decreasing $\mu$}.

\section{Radiation flux of the vaidya black hole}

\begin{figure*}
    \centering
    \subfloat{\includegraphics[width=.32\textwidth]{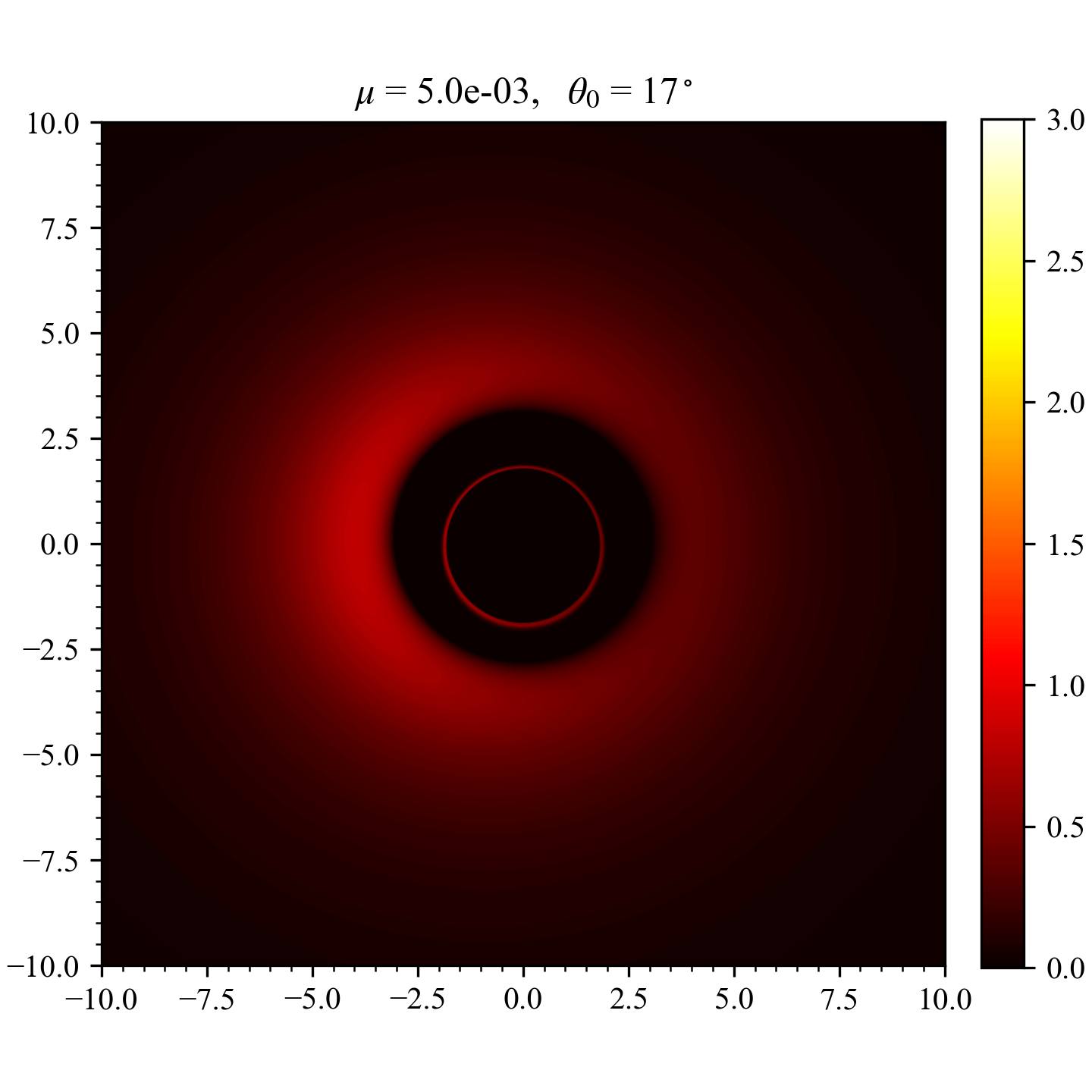}}
    \subfloat{\includegraphics[width=.32\textwidth]{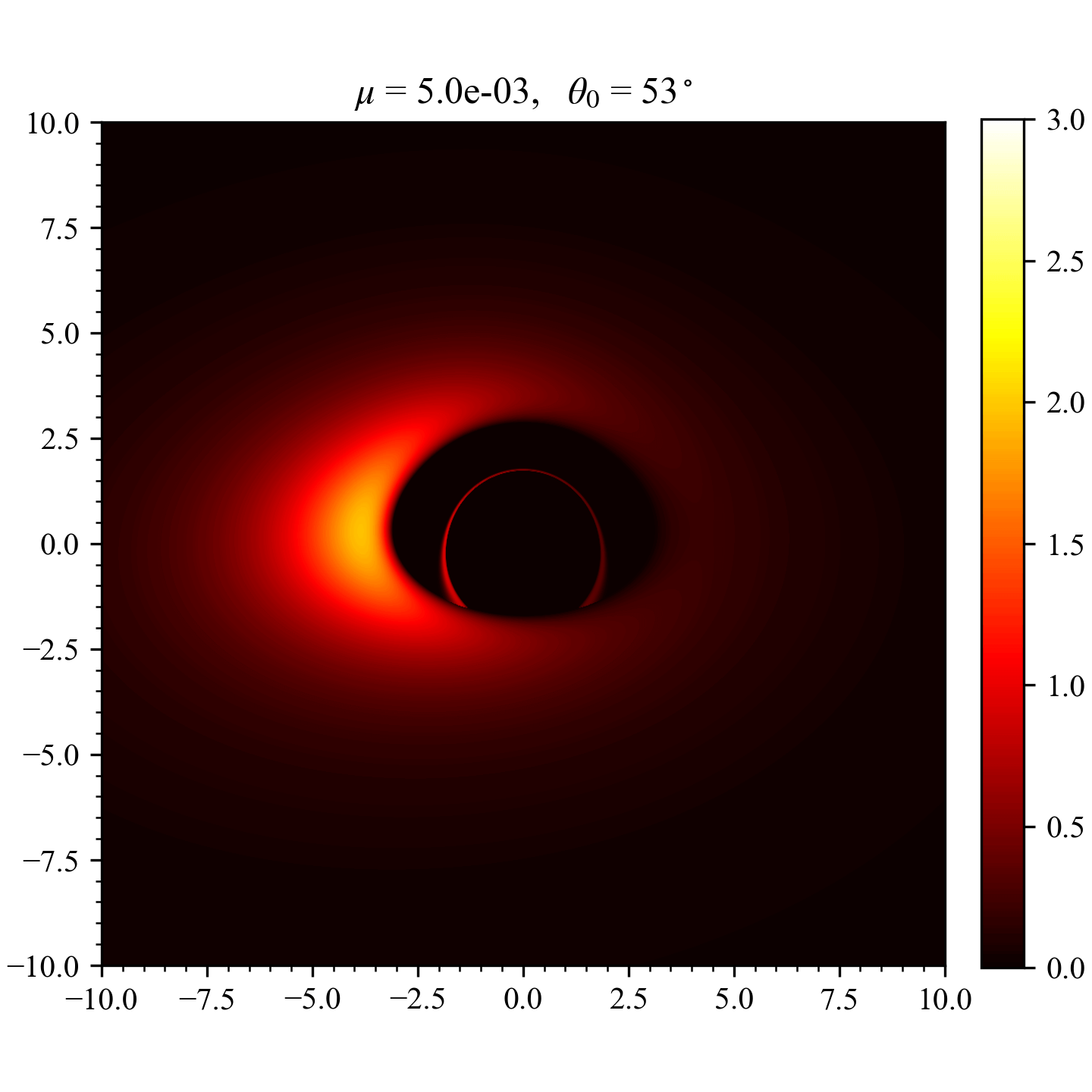}}
    \subfloat{\includegraphics[width=.32\textwidth]{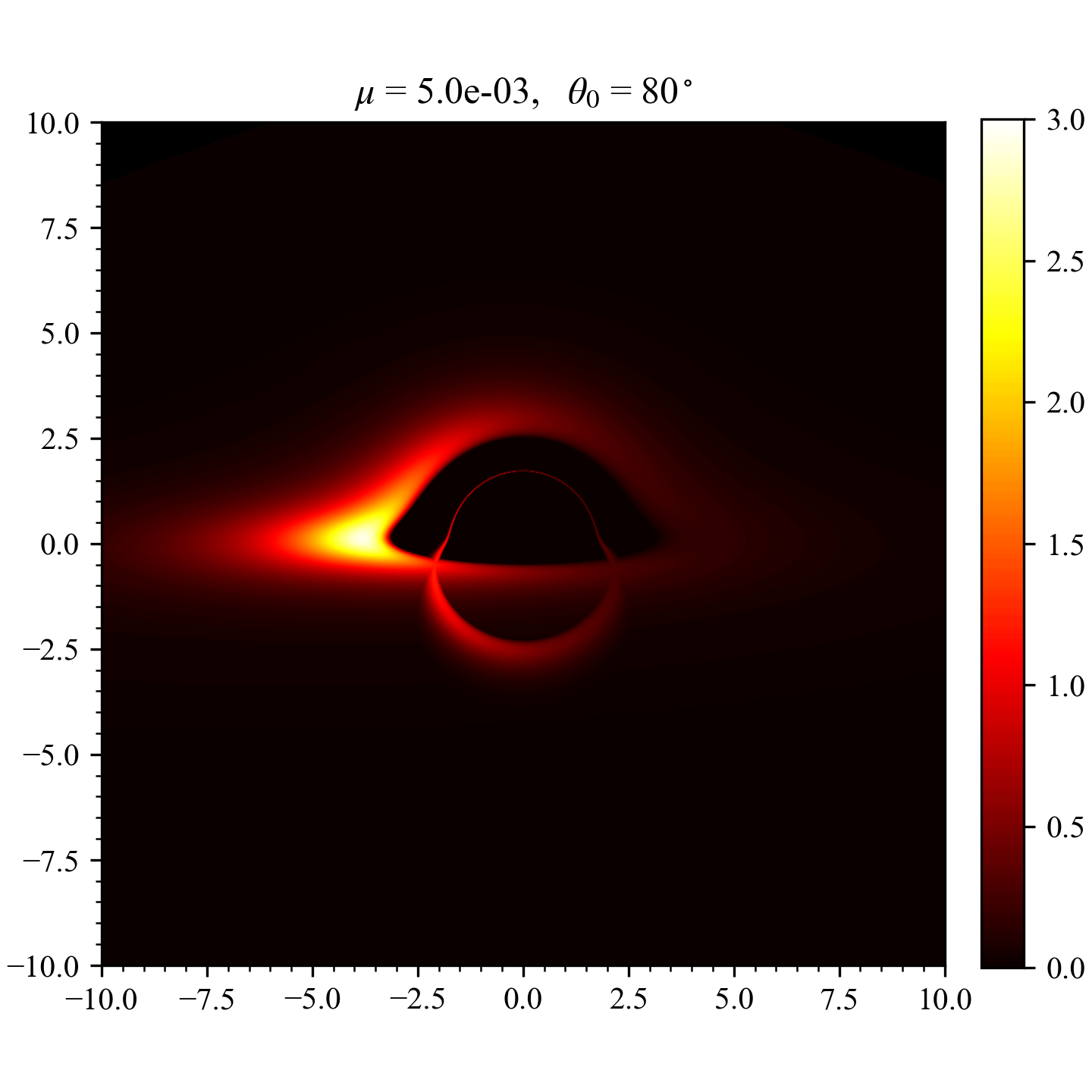}} \\
    \subfloat{\includegraphics[width=.32\textwidth]{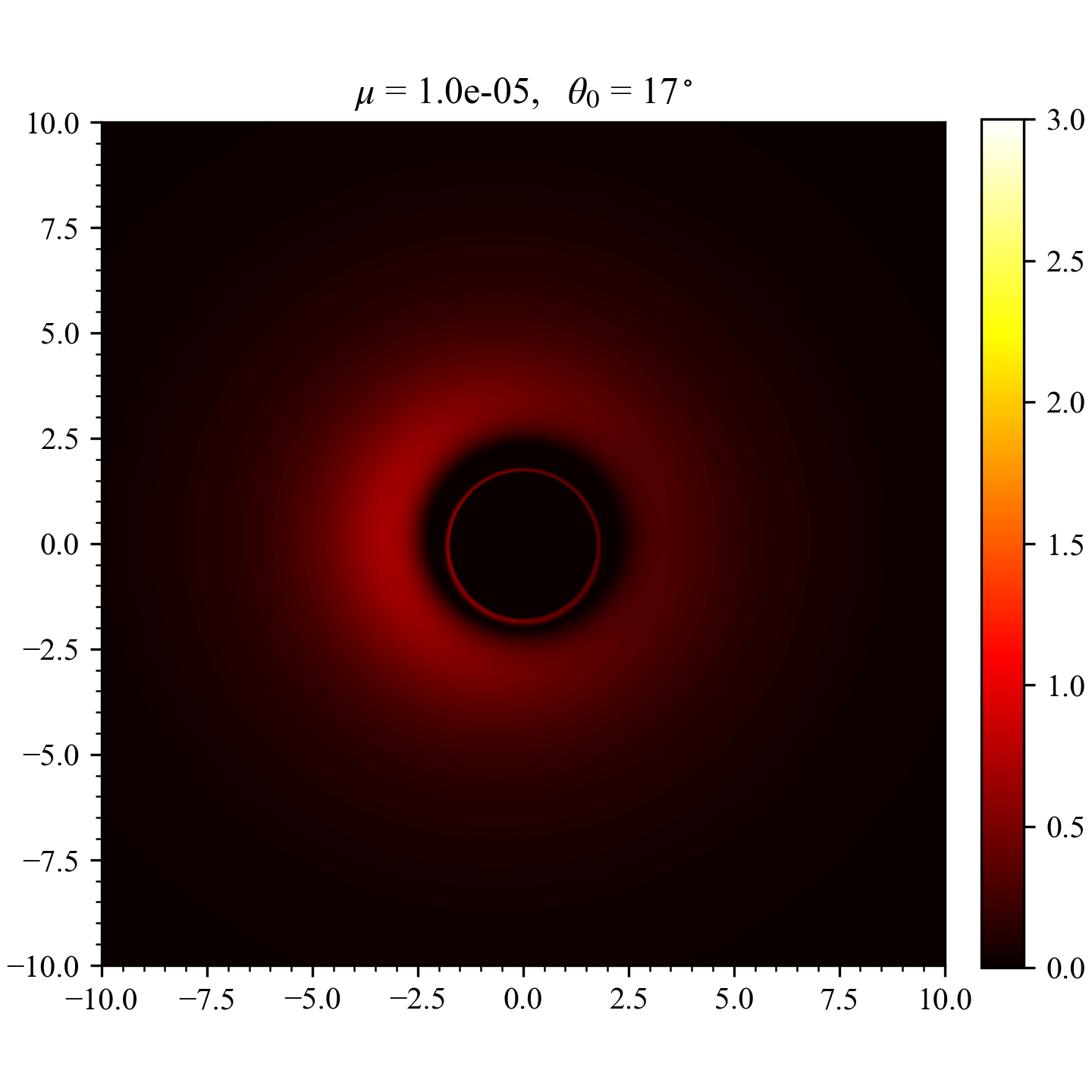}}
    \subfloat{\includegraphics[width=.32\textwidth]{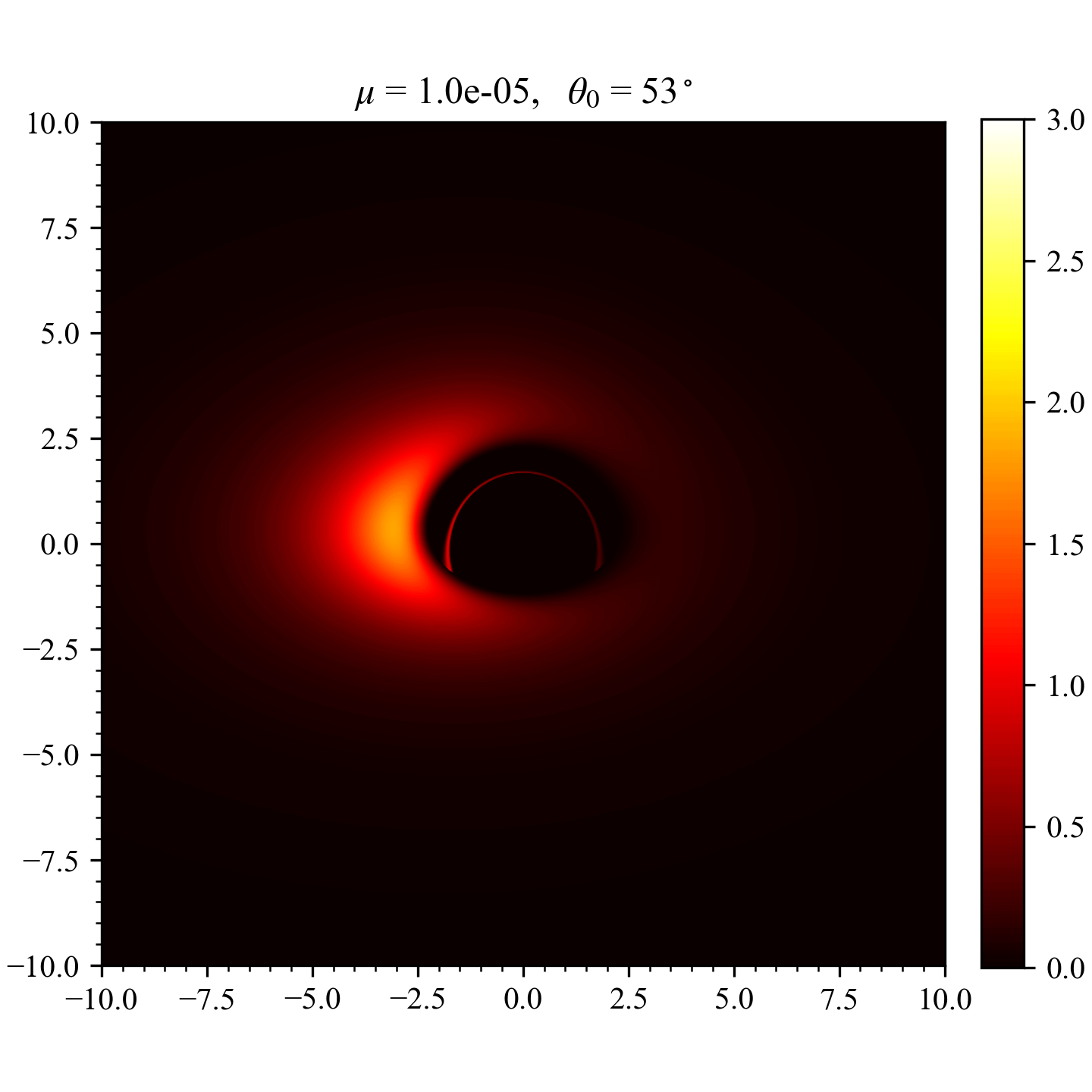}}
    \subfloat{\includegraphics[width=.32\textwidth]{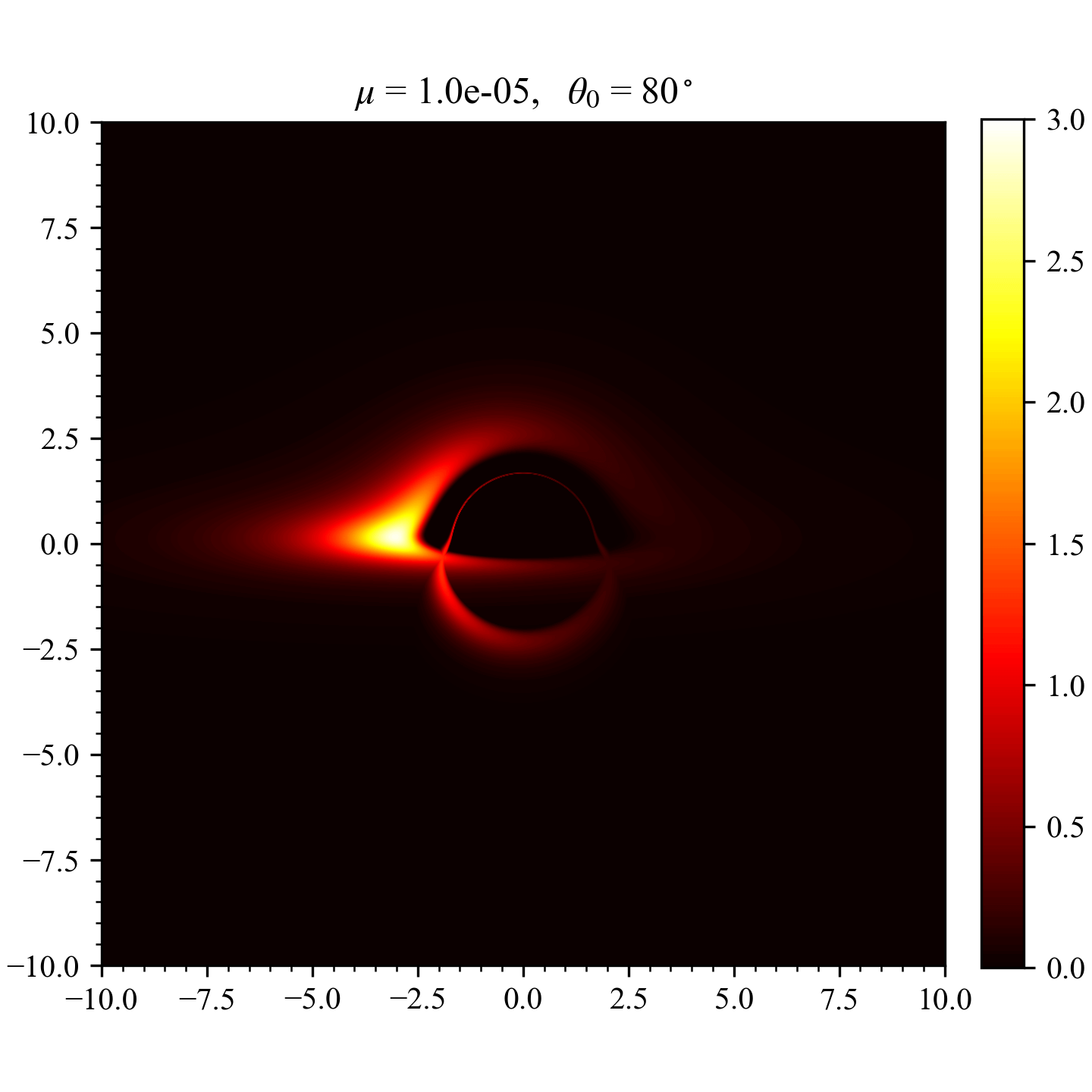}}
    \caption{\label{fig:c_flux}The observed flux \textcolor{black}{from} the conformal Vaidya black hole, surrounded by thin accretion disk.}
\end{figure*}

Previous studies have \textcolor{black}{derived} expressions for the radiative flux \textcolor{black}{in} thin accretion disk models, \textcolor{black}{where} the self-gravity \textcolor{black}{of disk} is neglected and the background spacetime is assumed to be stationary and asymptotically flat \citep{NovikovThorne1973,Page:1974he}. 
As shown earlier, the metric (\ref{eq:c_ds2}) is conformal to a static spacetime, and it can be readily verified that it becomes \textcolor{black}{asymptotically} flat at $R_+$. 
\textcolor{black}{Therefore}, the \textcolor{black}{radiative} flux in the conformal Vaidya background is given by the following expression
\begin{equation} \label{eq:intensity}
    I = - \frac{\dot{M_0}}{4\pi\sqrt{-g}} \frac{\Omega_{,R}}{(E-\Omega L)^2} \int_{R_{in}}^R (E-\Omega L) L_{,R} dR
\end{equation}
where $\dot{M_0}$ is the mass accretion rate, $g$ is the metric determinant, and $R_{in}$ is the inner boundary of the accretion disk, usually ISCO. 
$E$, $L$, and $\Omega$ represent the energy, angular momentum, and angular velocity, and are given by
\begin{equation}
    E = - \frac{g_{tt}}{\sqrt{-g_{tt} - g_{\phi\phi}\Omega^2}}
\end{equation}
\begin{equation}
    L = \frac{g_{\phi\phi} \Omega}{\sqrt{-g_{tt} - g_{\phi\phi}\Omega^2}}
\end{equation}
\begin{equation}
    \Omega = \frac{d\phi}{dt} = \sqrt{-\frac{g_{tt,R}}{g_{\phi\phi,R}}}
\end{equation}
For a thin accretion disk confined to the equatorial plane, the corresponding components are given by
\begin{equation}
    g_{tt} = - \Theta^2 f_c, \quad g_{\phi\phi} = \Theta^2 R^2, \quad g = - \Theta^8 R^4.
\end{equation}
By combining the above equations and \textcolor{black}{rearranging} Eq.~(\ref{eq:intensity}) to isolate the time variable $t_c$, we obtain
\begin{equation} \label{eq:intensity_0}
    I(t_c, R) = I(0, R) e^{-4 t_c} \equiv I_0(R) e^{-4 t_c}
\end{equation}
where $I_0(R)$ is the radiation flux at $t_c = 0$.

\textcolor{black}{Owing} to the gravitational \textcolor{black}{field} of the black hole, the observed flux \textcolor{black}{experiences} varying degrees of gravitational redshift. According to Ref.~\citep{Luminet:1979nyg}, \textcolor{black}{the observed flux can be expressed as}
\begin{equation} \label{eq:I_obs}
    I_{obs} = \frac{I}{(1+z)^4}
\end{equation}
where $(1+z)$ \textcolor{black}{denotes} the redshift factor, given by
\begin{equation}
    1+z = \frac{1 + b \Omega \sin \theta_0 \sin \alpha}{\sqrt{- g_{tt} - g_{\phi\phi} \Omega^2}}
\end{equation}
Similarly, let the redshift factor \textcolor{black}{be} $1 + z_0$ \textcolor{black}{at} $t_c = 0$, \textcolor{black}{then we obtain}
\begin{equation} \label{eq:zadd1}
    1 + z = (1 + z_0) e^{-t_c}
\end{equation}
By combining Eqs.~(\ref{eq:intensity_0}), (\ref{eq:I_obs}), and (\ref{eq:zadd1}), we find that the observed flux is independent of \textcolor{black}{the} time \textcolor{black}{coordinate} $t_c$, and can be expressed as
\begin{equation}
    I_{obs} = \frac{I_0}{(1 + z_0)^4}
\end{equation}

\begin{figure}
    \centering
    \includegraphics[width=.4\textwidth]{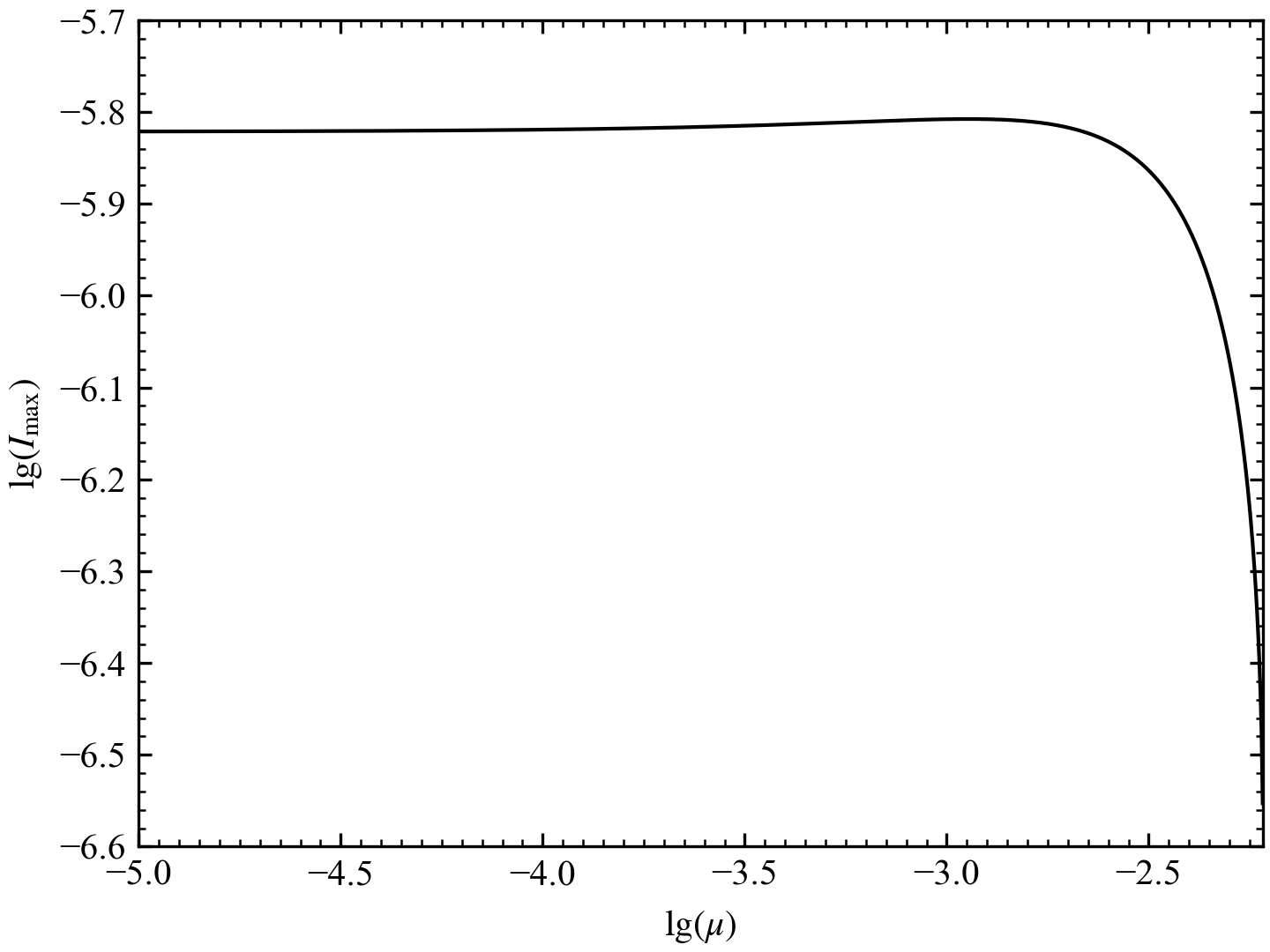}
    \caption{\label{fig:Imax_mu}The relationship between the peak radiation flux $I_{\max}$ and the parameter $\mu$ is illustrated, with both \textcolor{black}{quantities plotted} on a logarithmic scale.}
\end{figure}

Figure \ref{fig:c_flux} displays the observed flux from thin accretion disk in a conformal Vaidya spacetime for $\mu = 5 \times 10^{-3}$ and $10^{-5}$. 
The maximum flux value $I_{\max}$ is excluded in each panel, and the flux at each grid point \textcolor{black}{corresponds to} the sum of the direct and secondary observed fluxes. 
For \textcolor{black}{both} values of $\mu$, the distance between the observation plane and the black hole center is uniformly set to $0.9 R_+$. 
It can be seen that variation\textcolor{black}{s} in $\mu$ \textcolor{black}{have} a relatively small effect on the magnitude of the relative observed flux. 
Figure \ref{fig:Imax_mu} \textcolor{black}{depicts} the variation of the maximum observed flux $I_{\max}$ \textcolor{black}{as a function of} $\mu$. 
\textcolor{black}{As shown in the figure, $I_{\max}$ initially increases slightly and then decreases rapidly as $\mu$ increases.}

\section{Results}

\begin{figure}
    \centering
    \includegraphics[width=.4\textwidth]{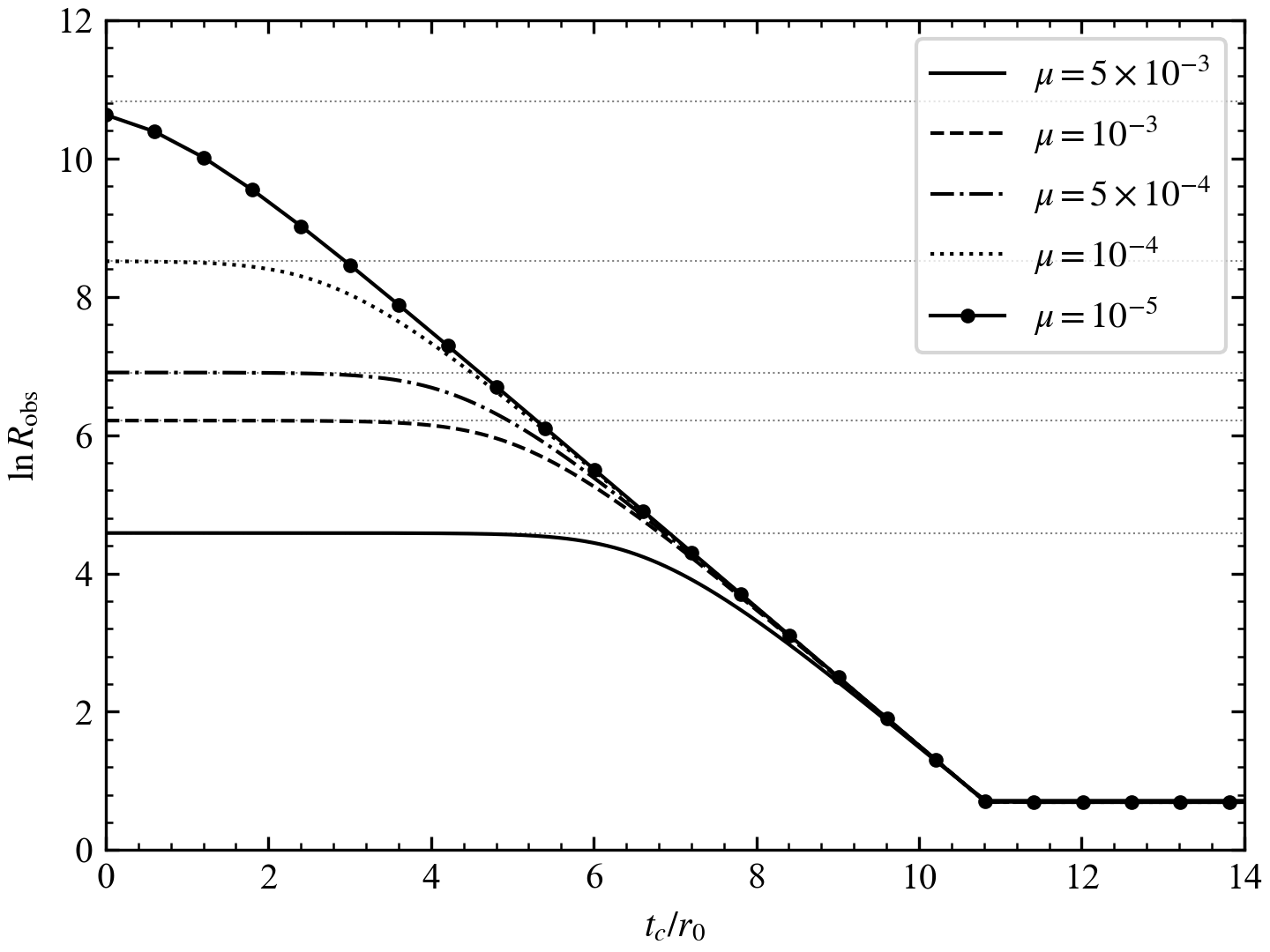}
    \caption{\label{fig:rcobs_tc}The \textcolor{black}{evolution} of $R_{obs}$ \textcolor{black}{as a function of} $t_c / r_0$ \textcolor{black}{is shown for an} observer \textcolor{black}{located at} $r_{obs} = 10^{5}$. The solid, dashed, dotted, dash-dotted, and solid-dot \textcolor{black}{curves} correspond to $\mu$ values of $5\times 10^{-3}$, $10^{-3}$, $5\times 10^{-4}$, $10^{-4}$, and $10^{-5}$, respectively. The gray dotted line \textcolor{black}{represents} the \textcolor{black}{conformal Killing horizon} $R_+$ \textcolor{black}{associated with each value of} $\mu$.}
\end{figure}

\begin{figure*}
    \centering
    \subfloat{\includegraphics[width=.32\textwidth]{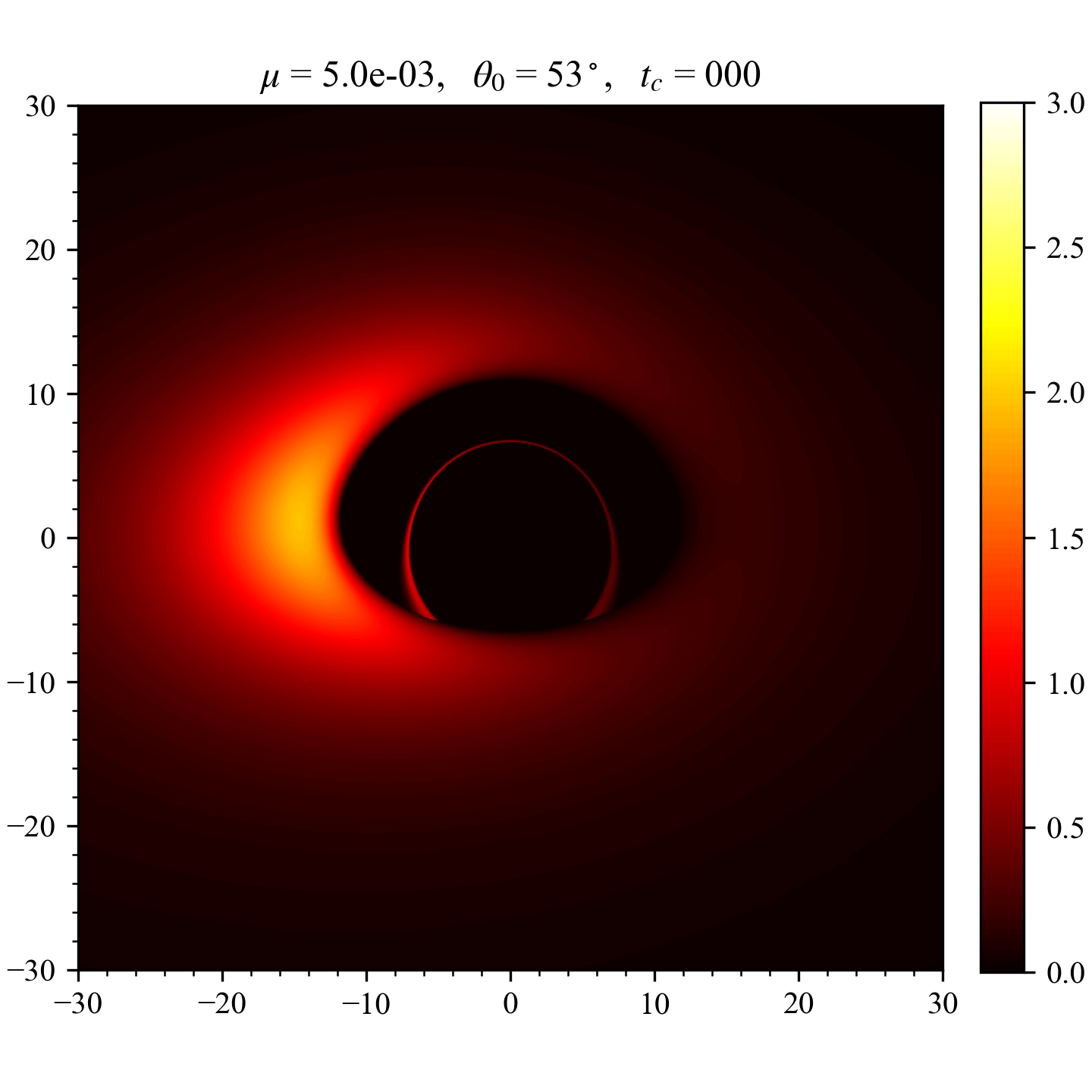}}
    \subfloat{\includegraphics[width=.32\textwidth]{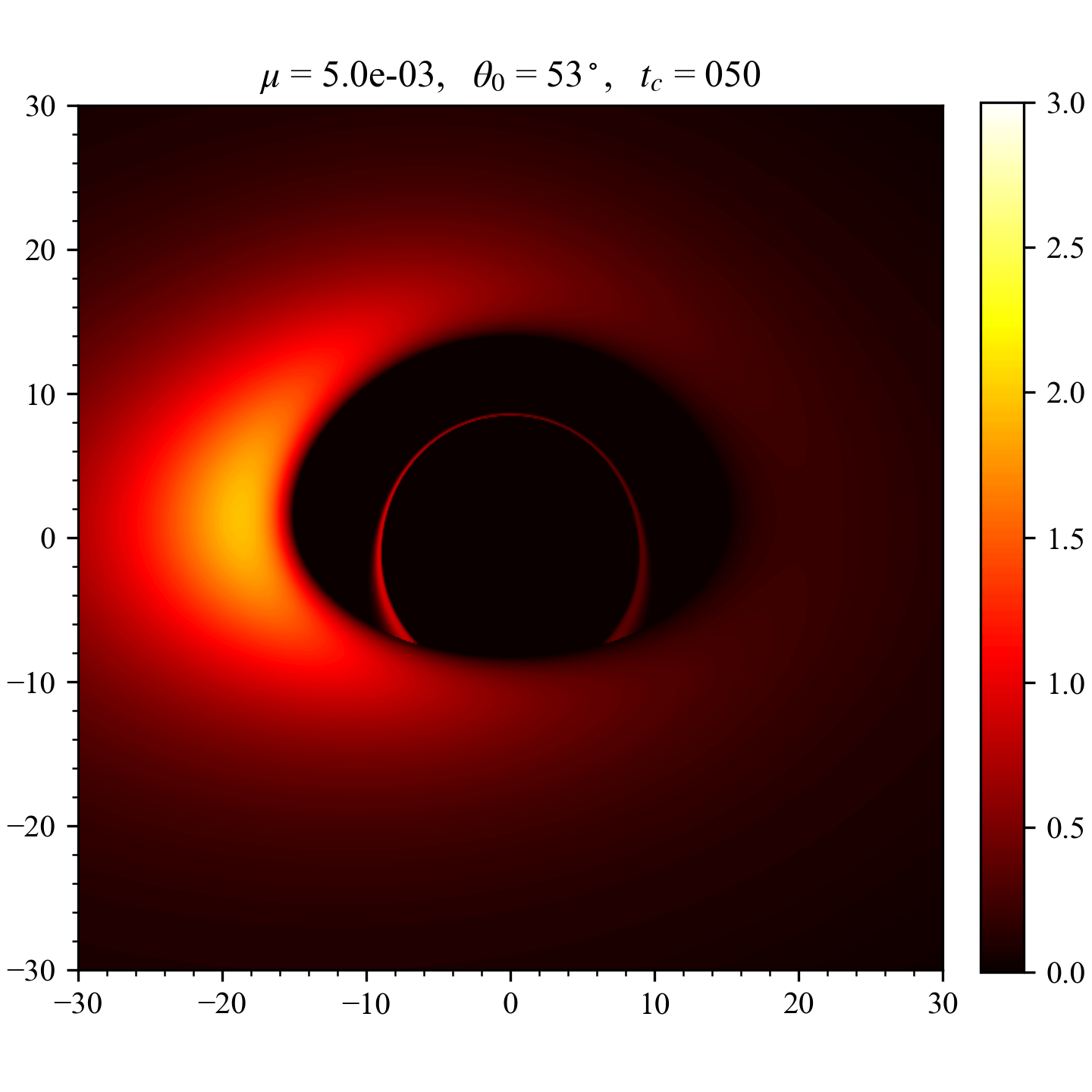}}
    \subfloat{\includegraphics[width=.32\textwidth]{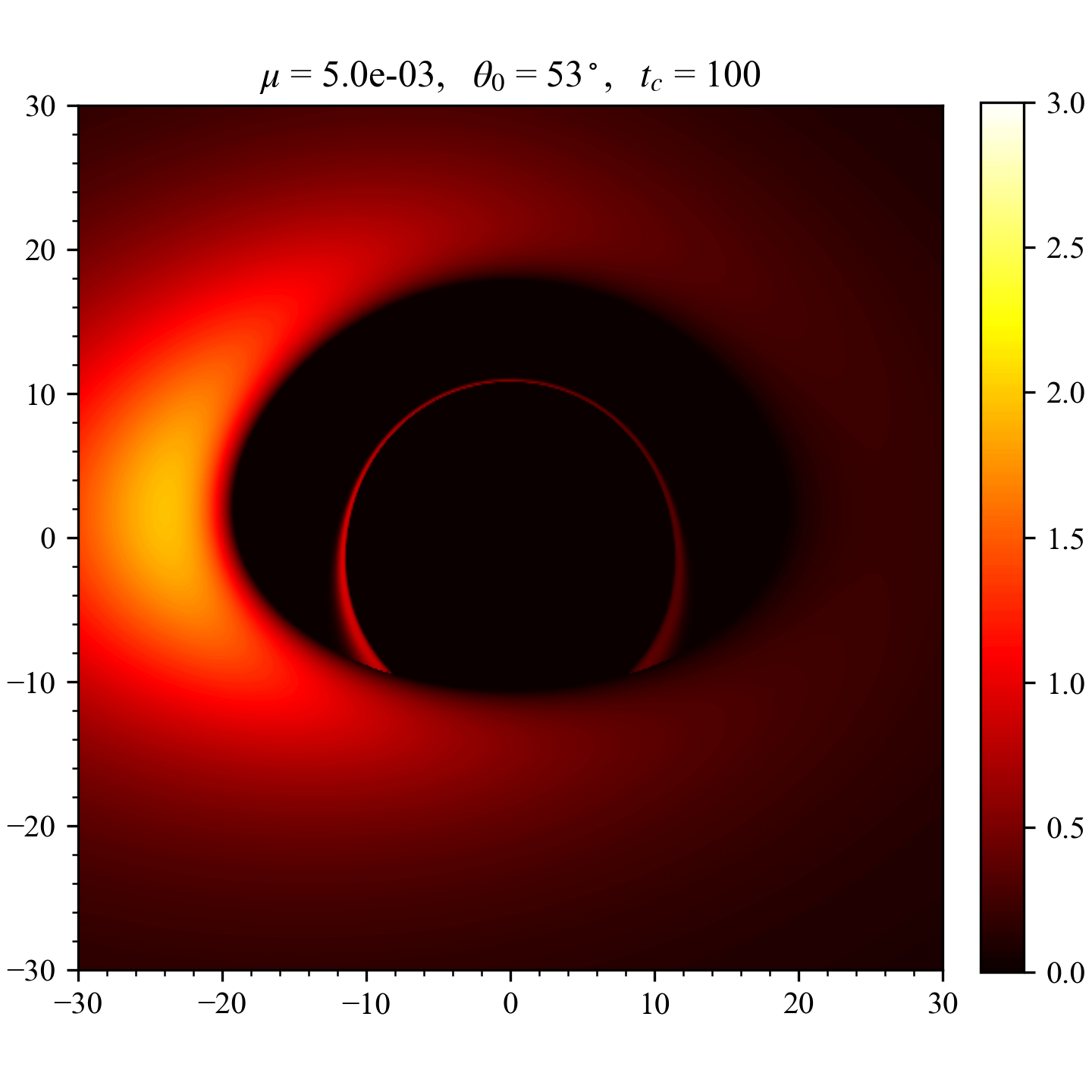}}\\
    \subfloat{\includegraphics[width=.32\textwidth]{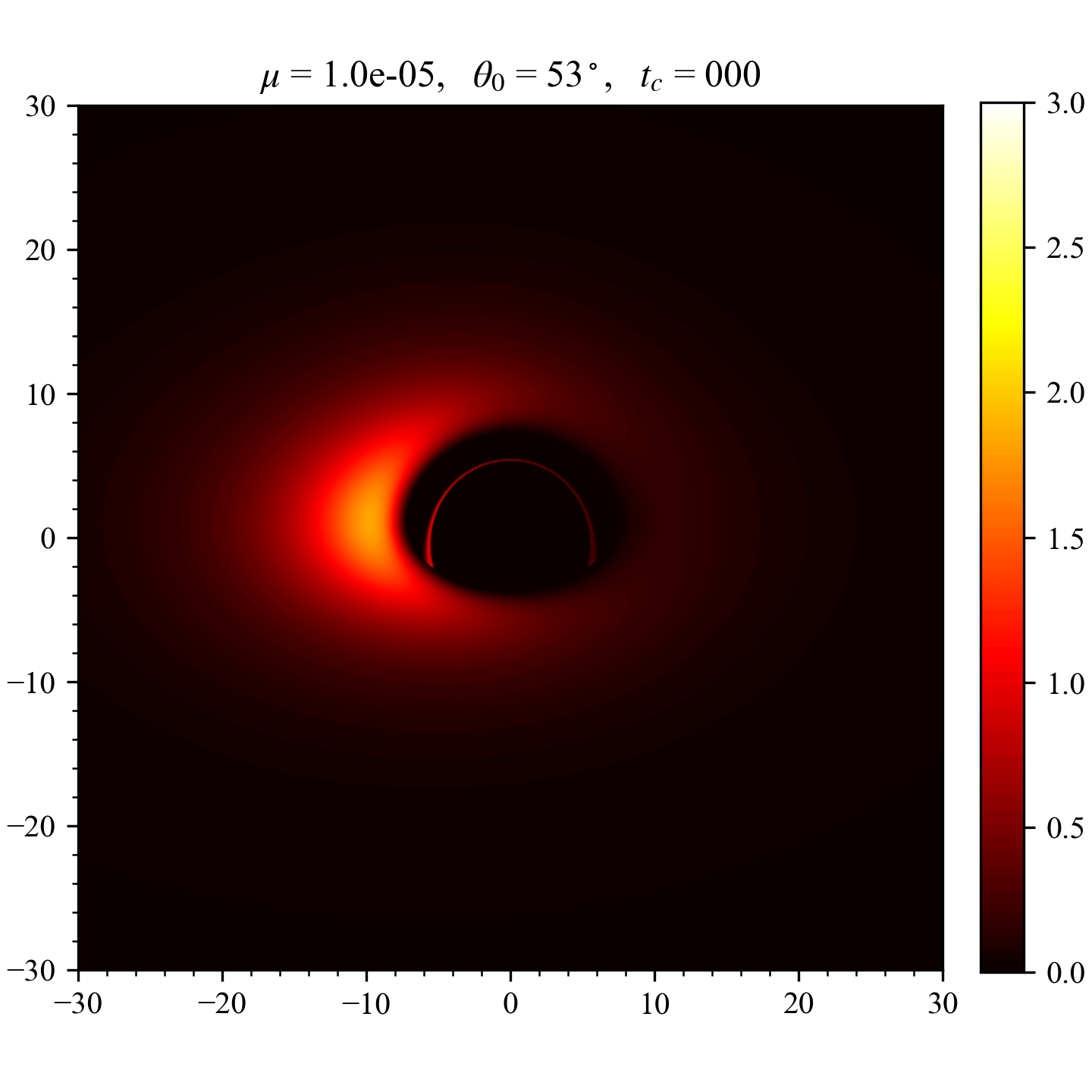}}
    \subfloat{\includegraphics[width=.32\textwidth]{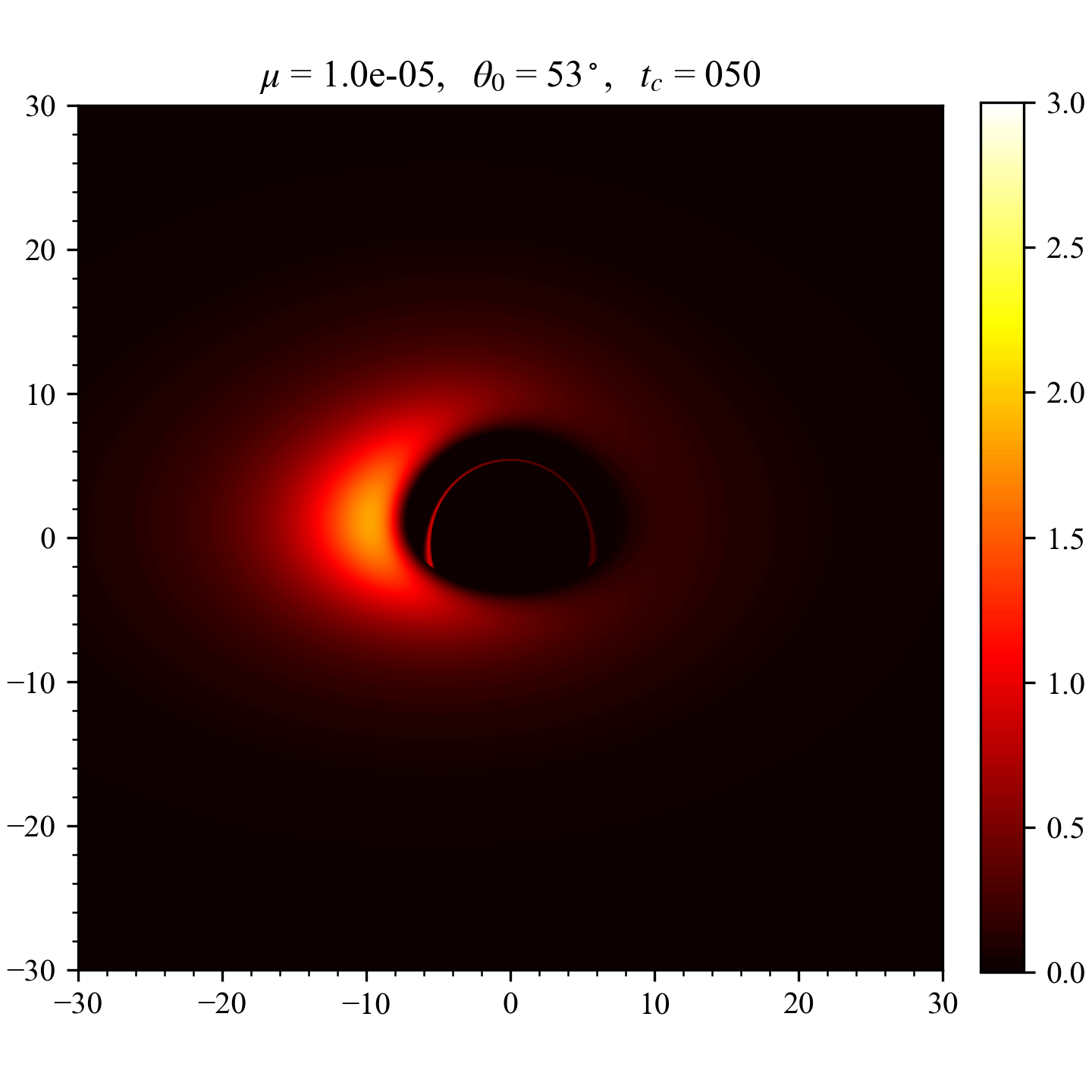}}
    \subfloat{\includegraphics[width=.32\textwidth]{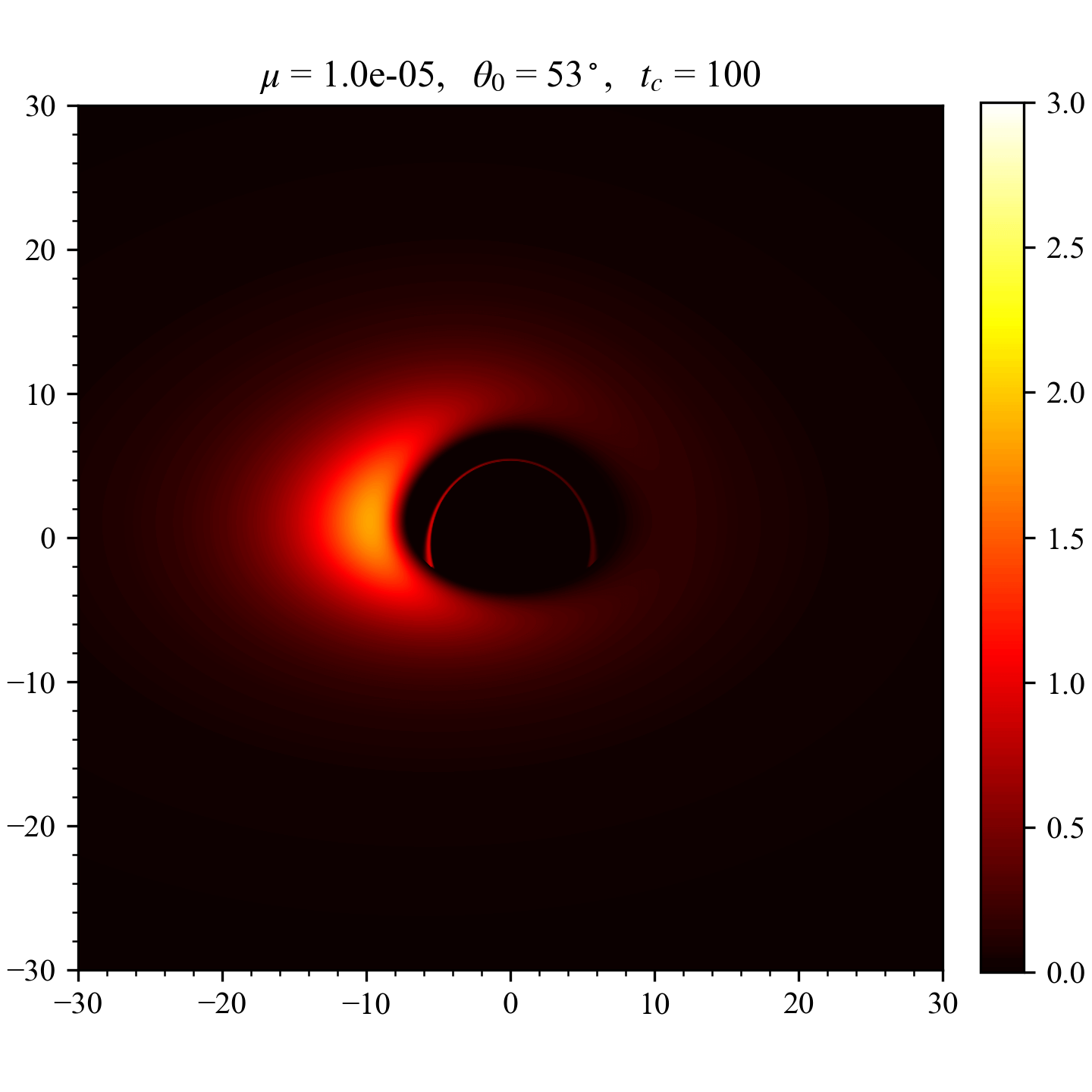}}
    \caption{\label{fig:flux}The observed flux of the Vaidya black hole \textcolor{black}{evolves} with \textcolor{black}{the time coordinate} $t_c$ for different values of $\mu$.}
\end{figure*}

\begin{figure}[htbp]
    \centering
    \includegraphics[width=.4\textwidth]{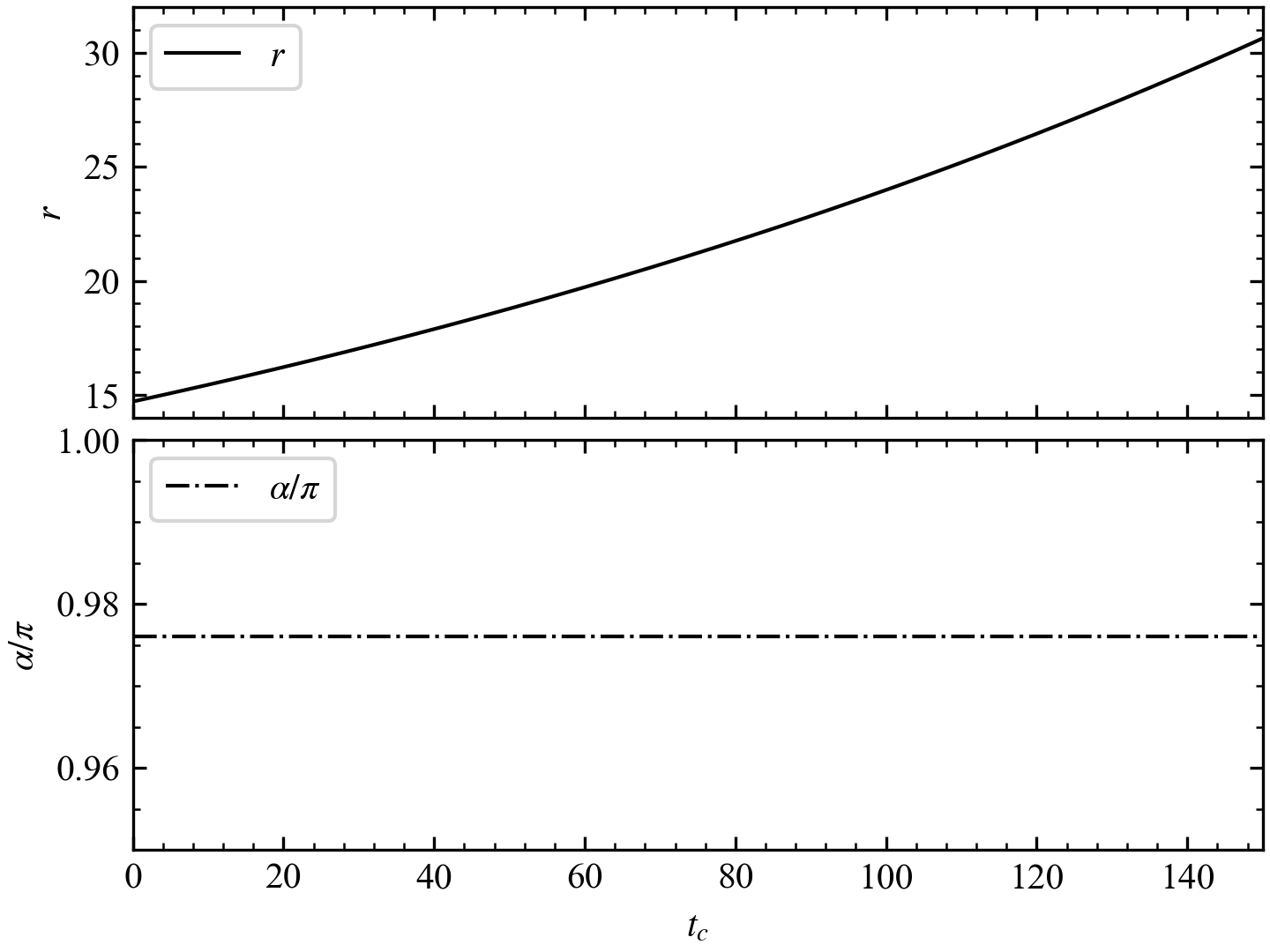}
    \caption{\label{fig:r_alpha_tc}The \textcolor{black}{evolution} of the observed flux peak position \textcolor{black}{as a function of} $t_c$ for $\mu = 5 \times 10^{-3}$. The upper \textcolor{black}{panel} shows the radial evolution, while the lower \textcolor{black}{panel} illustrates the angular \textcolor{black}{variation}. It \textcolor{black}{is} observed that the radial distance increases with \textcolor{black}{$t_c$}, \textcolor{black}{whereas} the angular position remains \textcolor{black}{nearly constant}.}
\end{figure}

By applying the conformal transformation \textcolor{black}{given} in Eq.~(\ref{eq:coord_trans}),  the coordinate system \textcolor{black}{can be mapped} from $(t_c, R, \theta, \phi)$ to $(t_c, r, \theta, \phi)$, \textcolor{black}{enabling} a more \textcolor{black}{physically intuitive} visualization of the Vaidya black hole image. 
This \textcolor{black}{transformation also facilitates further investigation into} how the image evolves with respect to the time parameter $t_c$. 
Although the time parameter $t_c$ \textcolor{black}{is} not be equivalent to the physical time $t$, the results still \textcolor{black}{provide} valuable insights into the appearance of a Vaidya black hole in \textcolor{black}{physical} spacetime.

Since \textcolor{black}{the coordinate transformation only modifies the radial component}, all angular quantities remain unchanged, although their \textcolor{black}{association} shifts from $R$ to $r$; the same applies to the radiation flux. 
Moreover, the observer’s position \textcolor{black}{is} fixed in the $(t_c, r, \theta, \phi)$ coordinate system, \textcolor{black}{implying} that $r_{obs}$ remains constant, while the corresponding \textcolor{black}{location} $R_{obs}$ in the conformal coordinates \textcolor{black}{evolves} dynamically  with $t_c$. 

Figure~\ref{fig:rcobs_tc} \textcolor{black}{illustrates the evolution of} $R_{obs}$ \textcolor{black}{as a function of} $t_c / r_0$ for \textcolor{black}{various} values of $\mu$, \textcolor{black}{with the observer fixed at} $r_{obs} = 10^{5}$. 
It is evident from Fig.~\ref{fig:rcobs_tc} that as $t_c$ increases, the increase in black hole mass is equivalent to the observer gradually approaching the black hole in the conformal coordinate system. 
Eventually, the observer \textcolor{black}{at} fixed $r_{obs}$ \textcolor{black}{asymptotically approaches} $R_-$ in the conformal \textcolor{black}{frame}. 
For small values of $\mu$, this \textcolor{black}{corresponding} time interval can be approximated as
\begin{equation}
    \Delta t_c \approx r_0 \ln \frac{r_{obs}}{R_-}
\end{equation}

Moreover, the projected distance $h_r$ on the observer’s plane in the $(t_c, r, \theta, \phi)$ coordinate system \textcolor{black}{can be derived} from Eqs.~(\ref{eq:hc}) and (\ref{eq:coord_trans}) \textcolor{black}{as follows}
\begin{equation}
    h_r = r_{obs} \tan{\alpha_{sh}} = h_c \Theta(t_c, R_{obs})
\end{equation}
With these preparations \textcolor{black}{in place}, the image of the Vaidya black hole \textcolor{black}{can now be constructed} in the $(t_c, r, \theta, \phi)$ coordinate system.

Figure \ref{fig:flux} \textcolor{black}{presents} the observed flux \textcolor{black}{at} different \textcolor{black}{values of} $t_c$ \textcolor{black}{for} $\mu = 5 \times 10^{-3}$ and $10^{-5}$. 
As \textcolor{black}{shown in Fig.}~\ref{fig:flux}, \textcolor{black}{for} $\mu = 5 \times 10^{-3}$, the shadow size increases significantly with \textcolor{black}{increasing} $t_c$, and the peak flux position shifts outward \textcolor{black}{in} the radial direction. 
\textcolor{black}{In contrast, for} $\mu = 10^{-5}$, the image \textcolor{black}{shows much weaker temporal variations} over the same interval. 
Figure \ref{fig:r_alpha_tc} \textcolor{black}{depicts} the evolution of the location of the maximum observed flux \textcolor{black}{as a function of} $t_c$ \textcolor{black}{for} $\mu = 5 \times 10^{-3}$. 
This further \textcolor{black}{confirms} that the peak flux \textcolor{black}{location moves} outward \textcolor{black}{in the radial direction} with time, \textcolor{black}{while remaining stationary} in the angular direction.

\section{Conclusions and discussion}

This paper investigates a time-dependent black hole described by the Vaidya metric, where the mass function $m(v)$ is assumed to increase linearly. 
\textcolor{black}{By introducing} a conformal factor and a redefined time coordinate \textcolor{black}{as} $t_c$, we analytically derived and numerically computed the ray trajectories and radiation flux in the conformal coordinate system. 
We further employed the conformal transformation to \textcolor{black}{explore} the time-dependent \textcolor{black}{behavior} of the Vaidya black hole in the $t_c - r$ coordinate system. Our \textcolor{black}{main findings can be summarized as follows}:

1. For an observer \textcolor{black}{located} at $r_{obs}$, the time-dependent \textcolor{black}{increase in} black hole mass \textcolor{black}{is} equivalent to the observer \textcolor{black}{gradually approaching} the black hole in the conformal coordinate system.

2. As \textcolor{black}{$t_c$} increases, the \textcolor{black}{apparent} size of the black hole shadow on the observation plane \textcolor{black}{grows accordingly}.

3. The maximum observed flux \textcolor{black}{initially increases slowly} and then \textcolor{black}{decreases rapidly as $\mu$ continues to grow}. 

4. \textcolor{black}{For a fixed value of $\mu$}, the facula on the observable plane \textcolor{black}{exhibits a radial shift over time, while its azimuthal position remains unchanged}. 
Since $\mu$ is typically small, \textcolor{black}{substantial radial displacement} may not be observable within a finite observational \textcolor{black}{timescale}. 
\textcolor{black}{However}, \textcolor{black}{such displacement may become detectable over a sufficiently long period}. 
The position-angle rotation of brightness asymmetry observed by the EHT may \textcolor{black}{instead be attributed to} variations in the emission source.

In future \textcolor{black}{work}, we aim to \textcolor{black}{provide a theoretical explanation for} and reproduce the counterclockwise rotation of the brightest region observed by the EHT. 
To this end, we \textcolor{black}{plan to investigate} more \textcolor{black}{sophisticated} time-dependent black hole metrics and \textcolor{black}{models of} radiation source.

\section*{Acknowledgments}

This work is supported by the Brazilian agencies Fapesq-PB, Fund Project of Chongqing Normal University (Grant Number: 24XLB033) and the National Natural Science Foundation of China (\textcolor{black}{Grants Number:} 42230207, 42074191). 




\bibliographystyle{elsarticle-harv} 
\bibliography{vaidya}






\end{document}